\documentclass[
reprint,
superscriptaddress,
amsmath,amssymb,
aps,
prb,
floatfix
]{revtex4-2}

\usepackage{graphicx}
\usepackage{dcolumn}
\usepackage{bm}
\usepackage[colorlinks,citecolor=magenta]{hyperref}

\begin{document}

\title{Coupling of MoS$_2$ Excitons with Lattice Phonons and Cavity Vibrational Phonons in Hybrid Nanobeam Cavities}

\author{Chenjiang Qian}
\email{chenjiang.qian@wsi.tum.de}
\author{Viviana Villafañe}
\affiliation{Walter Schottky Institut and Physik Department, Technische Universit{\" a}t M{\" u}nchen, Am Coulombwall 4, 85748 Garching, Germany}
\author{Marko M. Petrić}
\affiliation{Walter Schottky Institut and Department of Electrical and Computer Engineering, Technische Universit{\" a}t M{\" u}nchen, Am Coulombwall 4, 85748 Garching, Germany}
\author{Pedro Soubelet}
\author{Andreas V. Stier}
\author{Jonathan J. Finley}
\email{finley@wsi.tum.de}
\affiliation{Walter Schottky Institut and Physik Department, Technische Universit{\" a}t M{\" u}nchen, Am Coulombwall 4, 85748 Garching, Germany}

\begin{abstract}
	We report resonant Raman spectroscopy of neutral excitons X$^0$ and intravalley trions X$^-$ in hBN-encapsulated MoS$_2$ monolayer embedded in a nanobeam cavity.
	By temperature tuning the detuning between Raman modes of MoS$_2$ lattice phonons and X$^0$/X$^-$ emission peaks, we probe the mutual coupling of excitons, lattice phonons and cavity vibrational phonons.
	We observe an enhancement of X$^0$-induced Raman scattering and a suppression for X$^-$-induced, and explain our findings as arising from the tripartite exciton-phonon-phonon coupling.
	The cavity vibrational phonons provide intermediate replica states of X$^0$ for resonance conditions in the scattering of lattice phonons, thus enhancing the Raman intensity.
	In contrast, the tripartite coupling involving X$^-$ is found to be much weaker, an observation explained by the geometry-dependent polarity of the electron and hole deformation potentials.
	Our results indicate that phononic hybridization between lattice and nanomechanical modes plays a key role in the excitonic photophysics and light-matter interaction in 2D-material nanophotonic systems.
\end{abstract}

\maketitle

Nano-opto-electro-mechanical systems are of strong interest in the study of light-matter interactions since they intentionally couple electronic, optical and vibrational degrees of freedom having vastly different eigenfrequencies \cite{Mahboob2012,RevModPhys.86.1391,PhysRevX.8.021052,Midolo2018}.
Hereby, interband optical response becomes sensitive to the local optical field and the state of motion (phonons) in the nanosystem \cite{Wei2020,Zhang2022}.
Monolayer transition metal dichalcogenides (TMDs) are of particular interest in this context since they (i) can be attached via van der Waals bonding to a wide range of different substrates, and they combine (ii) strong light-matter interactions through excitonic transitions at room temperature with (iii) large photoelastic coupling strengths to the local deformations
\cite{Ghorbani-Asl2013,Miller2019,doi:10.1021/acs.nanolett.6b02038,C9NR02447F}.
Recent works on 2D-material nanophotonic cavities report that phonons modulate the light-matter interaction by limiting the exciton mobility \cite{PhysRevLett.128.237403} and introducing vibronic sublevels \cite{PhysRevLett.126.227401,PhysRevLett.128.087401}.
These phonon-mediated effects indicate wide potentials of phononic technology in the cavity QED study.

While these recent works spell out the key role played by phonons in 2D-material nanocavities \cite{PhysRevLett.128.237403,PhysRevLett.126.227401,PhysRevLett.128.087401,Rosser2020}, the additional degree of freedom associated with the phononic vibration from cavity nanomechanical modes and the delicate interplay between lattice phonons, nanomechanical modes, excitons, and cavity photons have not been previously elucidated.
Indeed, besides the lattice phonons from atomic vibrational modes \cite{Selig2016,PhysRevLett.116.127402,PhysRevMaterials.2.054001,PhysRevB.99.085412}, there also exist the phononic vibrational modes of nanocavities \cite{10.1038/nature08061,10.1038/nature08524,Zalalutdinov2021,doi:10.1021/acs.nanolett.0c05089}.
The cavity vibrational phonons also introduce deformations and interact with excitons \cite{doi:10.1021/nl501413t,Chen2015}.
Therefore, understanding the interplay between different phonons and their mutual coupling to excitons are the key to further explore and control phonon-mediated processes in light-matter interactions.

In this letter, we reveal the tripartite coupling between excitons, lattice phonons and cavity vibrational phonons in the cavity-MoS$_2$ system using resonant Raman spectroscopy.
We use optimized high-Q nanobeam cavities to embed hBN-encapsulated monolayer MoS$_2$ \cite{PhysRevLett.128.237403}.
The encapsulation suppresses disorder-induced fluctuations \cite{PhysRevX.7.021026,Wierzbowski2017,Raja2019} and allows clear spectral separation of \textit{pristine} neutral excitons X$^0$ and intravalley trions X$^-$.
We tune the Raman modes of MoS$_2$ lattice phonons through the exciton emission energies by the temperature \cite{doi:10.1063/1.4862859}.
The X$^0$(X$^-$)-phonon coupling strengths in the X$^0$(X$^-$)-induced Raman scattering are revealed by the X$^0$-Raman and X$^-$-Raman resonant peaks in the detuning dependent Raman intensity \cite{Molas2017,doi:10.1063/1.4862859,Carvalho2017,gontijo2019}.
We observe a significant enhancement of X$^0$-induced Raman scattering and a suppression for X$^-$-induced.
The enhancement of X$^0$-induced scattering is well explained by exciton-phonon-phonon coupling, where the cavity vibrational phonons provide intermediate replica states for resonance conditions in the scattering of lattice phonons.
In contrast, the X$^-$--cavity-phonon coupling is very weak, explained by the near cancellation of electron and hole deformation potentials in the ribbon-shaped MoS$_2$ monolayers \cite{doi:10.1021/ja4109787}.
Thereby, the intermediate state which enhances the Raman intensity does not occur for X$^-$.
The temperature dependence of Raman enhancement reveals that a discrete number of cavity phonons participate in the coupling, further supporting the phononic hybridization between material and nanomechanical degrees of freedom in the quantum system.

\begin{figure*}
	\includegraphics[width=\linewidth]{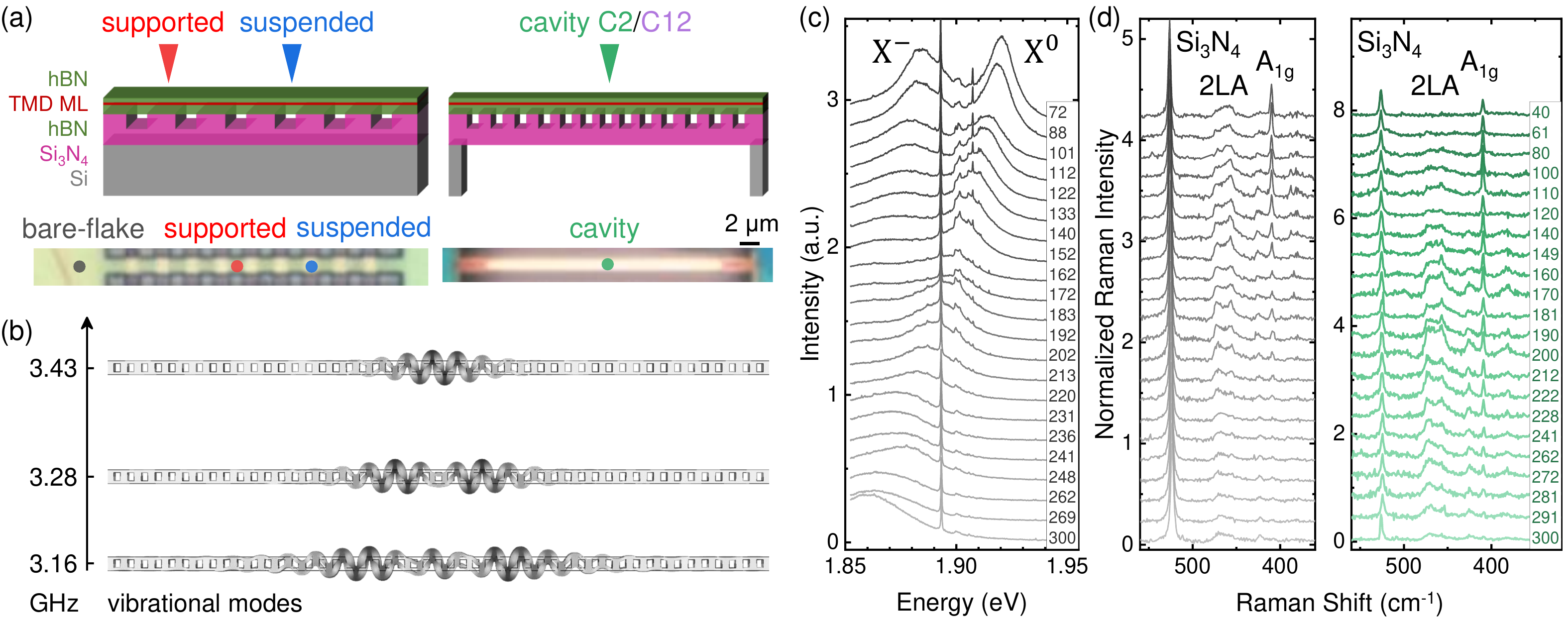}
	\caption{\label{f1}
		(a) Schematic of four distinct laser positions for spectroscopy: bare flake off to the side of nanobeam on the planar substrate, supported on Si$_3$N$_4$ and suspended on etched Si$_3$N$_4$ part of the nanobeam, and the center position of the nanobeam cavity.
		(b) Calculated vibrational modes of the nanobeam cavity.
		The MoS$_2$ deforms primarily along the axis of nanobeam.
		(c) Temperature-dependent raw spectra recorded from the bare flake showing both Raman and photoluminescence (PL) signals.
		(d) Raman spectra after subtracting the PL emission baseline recorded from the bare flake (gray) and cavity C2 (green).
		Intensities are normalized to the Si$_3$N$_4$ peak.
		Temperatures are denoted in Kelvin in the inset box in (c)(d).
	}
\end{figure*}

Our sample structures are depicted in Fig.~\ref{f1}(a).
The hBN/MoS$_2$/hBN heterostructure is prepared using mechanical exfoliation and viscoelastic dry transfer methods \cite{PhysRevLett.128.237403,Pizzocchero2016}.
The monolayer MoS$_2$ is encapsulated by the top (bottom) hBN with a thickness around 15 (55) nm, transferred onto a 200 nm thick Si$_3$N$_4$ layer on a Si substrate.
The sample is patterned into series of photonic crystal nanobeams \cite{PhysRevLett.128.237403,10.1021/acs.nanolett.2c00739}.
We investigate the Raman spectra recorded from four kinds of positions.
The first case bare flake corresponds to the region consisting of only hBN-encapsulated MoS$_2$ on the planar Si$_3$N$_4$ substrate.
Data recorded from this case are denoted by gray datasets in this work.
The second and third cases, denoted by red and blue datasets, correspond to supported and suspended positions in the nanobeam, respectively.
Since the photonic crystal trenches in this sample have a periodicity of $2\ \mathrm{\mu m}$ and a lateral size of $1\ \mathrm{\mu m}$, the laser spot can be precisely and readily positioned on the supported or suspended positions, respectively.
These three cases (bare flake, supported, suspended) are control experiments, in contrast to the fourth case cavity corresponding to the center position of high-Q cavities.
In this work, we present data recorded from two cavities with the different nanobeam width 520 (420) nm for cavity C2 (C12).
The data recorded from cavity C2 (C12) is denoted by green (purple) datasets, respectively.
Optical and vibrational modes of the cavity are formed by chirping the photonic crystal periodicity to create photonic and phononic band gap confinement \cite{PhysRevLett.128.237403,10.1038/nature08524}.
Typical vibrational modes calculated via fully 3D finite element simulation are presented in Fig.~\ref{f1}(b), and the calculation details are presented in Sec.~\ref{secs2} in Supplement \cite{supplement}.

The three control experiments, i.e., bare flake, supported and suspended cases are chosen to identify other factors besides the optical and vibrational modes which might affect the Raman properties in the cavity.
For example, in the cavity two types of local static strain are induced in the TMD: tensile strain from Si$_3$N$_4$ structures \cite{doi:10.1021/acs.nanolett.2c00613,Chai2017} and strain arising from the 2D heterostructure being freely suspended.
Effects from the former can be isolated by the supported case, and effects from the latter can be revealed by the suspended case.
The reactive ion etching during nanofabrication might also affect the hBN/TMD/hBN heterostructure \cite{Kim2018,doi:10.1002/adom.201801344}, and if this is the case, the effect on Raman properties can be revealed by both supported and suspended cases.

We implement resonant Raman spectroscopy by varying the lattice temperature to tune the Raman modes of MoS$_2$ lattice phonons through the X$^0$ and X$^-$ emission.
The excitation cw-laser laser has the wavelength 632 nm with a spot size $\sim 1\ \mathrm{\mu m}$ and power $\sim 100\ \mathrm{\mu W}$.
The excitation conditions produce both exciton emissions and Raman signals superimposed in spectra, as shown in the raw spectra measured from the bare flake in Fig.~\ref{f1}(c).
Raman spectra are then extracted by subtracting the emission baseline, and the results from the bare flake and cavity C2 are presented in Fig.~\ref{f1}(d).
We observe three dominant Raman modes: the Si$_3$N$_4$ phonon ($525\ \mathrm{cm^{-1}}$) and two MoS$_2$ lattice phonons as acoustic phonon 2LA ($450-480\ \mathrm{cm^{-1}}$) and optical phonon A$_{1g}$ ($409\ \mathrm{cm^{-1}}$) \cite{Saito_2016,PhysRevB.84.155413}.
The intensity of 2LA peak in both cases exhibits a clear maximum around the resonance to X$^0$, which is a typical resonant Raman phenomenon \cite{doi:10.1063/1.4862859,Molas2017,Carvalho2017}.
Meanwhile, differences are observed between cavities and control experiments, e.g., the intensity of 2LA peak at low temperature, indicating the phonons and exciton-phonon couplings are modulated in the cavity.

\begin{figure}
	\includegraphics[width=\linewidth]{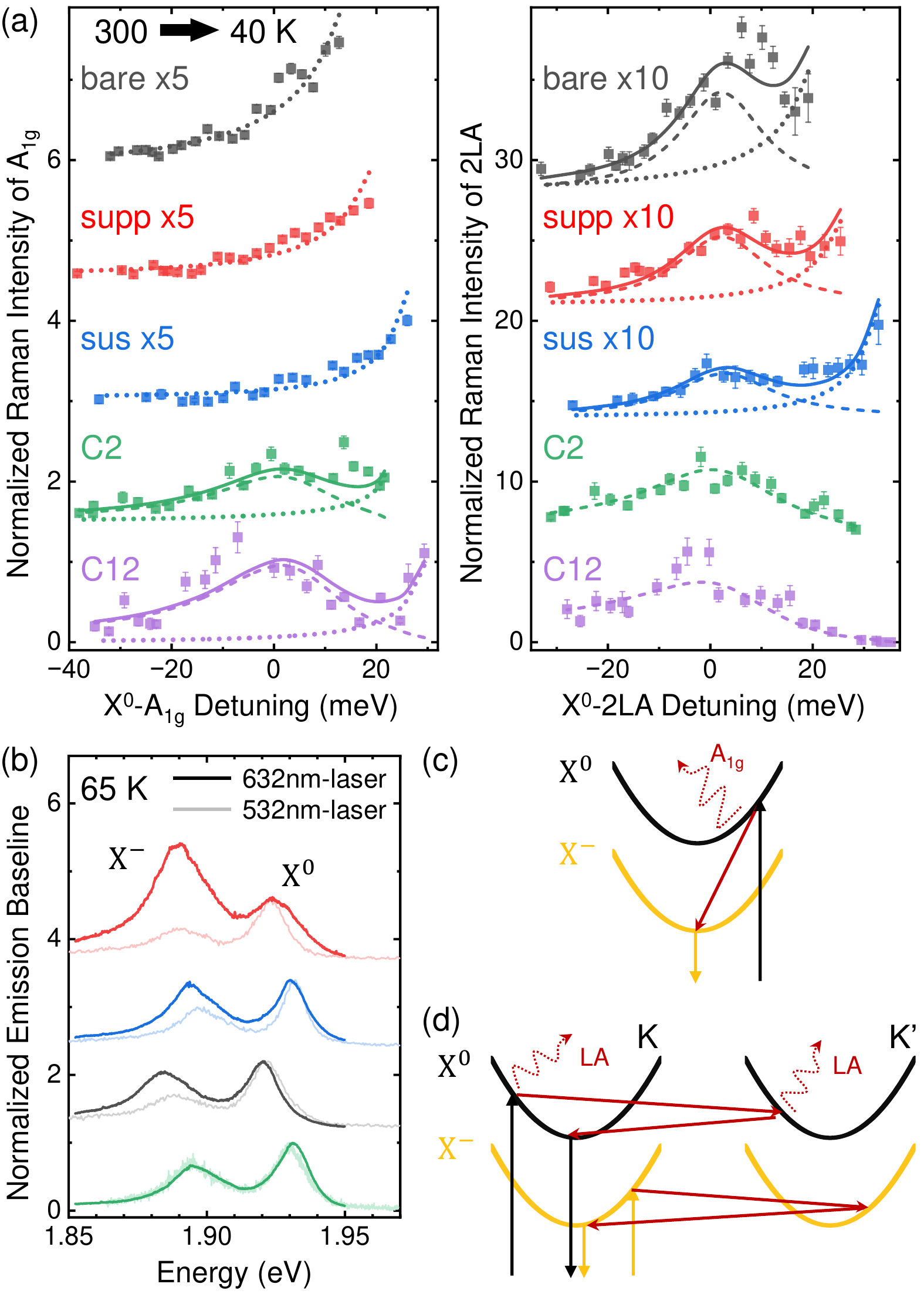}
	\caption{\label{f2}
		(a) Normalized Raman intensities of A$_{1g}$ and 2LA.
		Dashed (dotted) peaks are from the X$^0$(X$^-$)-phonon coupling.
		In control cases (gray, red, blue) the coupling strength $g_{X,p}$ is constant.
		In cavities (green, purple), $g_{\mathrm{X}^-,p}$ is constant and $g_{\mathrm{X}^0,p}$ is enhanced with a $T^N$ dependence.
		(b) Comparisons between exciton emission recorded using 632 nm-laser (dark) and 532 nm-laser (light) at 65 K, normalized by the X$^0$ peak.
		(c) Schematic of doubly resonant scattering of A$_{1g}$ around X$^-$-Raman resonance, which converts X$^0$ to X$^-$, thereby enhances the X$^-$ emission in (b) for the three control cases.
		(d) Schematic of the doubly resonant scattering of 2LA around X$^0$(X$^-$)-Raman resonance.
	}
\end{figure}

To further investigate the exciton-phonon couplings, we normalize the Raman intensities by dividing the integrated peak intensity by the Bose factor and the intensity of Si$_3$N$_4$ peak \cite{C9NR02447F}, i.e.
\begin{eqnarray}
	I'_{\mathrm{A}_{1g}}=\frac{I_{\mathrm{A}_{1g}}/(n_{\mathrm{A}_{1g}}+1)}{I_{\mathrm{SiN}}/(n_{\mathrm{SiN}}+1)},\
	I'_{\mathrm{2LA}}=\frac{I_{\mathrm{2LA}}/(n_{\mathrm{LA}}+1)^2}{I_{\mathrm{SiN}}/(n_{\mathrm{SiN}}+1)} \nonumber
\end{eqnarray}
where $I_{\mathrm{A}_{1g}}$, $I_{\mathrm{SiN}}$ and $I_{\mathrm{2LA}}$ are peak intensities extracted from Raman spectra.
$n_{\mathrm{A}_{1g}},\ n_{\mathrm{SiN}},\ n_{\mathrm{LA}}$ are temperature-dependent Bose distribution factors
\begin{eqnarray}
	n_p=\frac{1}{e^{{\hbar}\left(\omega_{\mathrm{Laser}}-\omega_p\right)/\left( k_\mathrm{B} T \right)}-1}, p=\{\mathrm{A}_{1g}, \mathrm{2LA}, \mathrm{SiN}\} \nonumber
\end{eqnarray}
where $\hbar\omega_{\mathrm{Laser}}$ ($\hbar\omega_{p}$) is the energy of the laser (Raman mode), $k_\mathrm{B}$ is the Boltzmann constant, and $T$ is the temperature.
Theoretically, the normalized Raman intensity consists of signals from all possible light-matter intermediate states, according to \cite{doi:10.1002/9783527632695.ch5}
\begin{eqnarray}
	\displaystyle\sum_{m}\left|\frac{M_{fm}M_{ep}M_{mi}}{\left(\omega_{m}-(i/2)\gamma_m-\omega_{\mathrm{Laser}}\right)\left(\omega_{m}-(i/2)\gamma_m-\omega_{p}\right)}\right|^2 \nonumber,
\end{eqnarray}
where $i,\ m$ and $f$ denote the initial, intermediate and final states, $\omega_{m}$ and $\gamma_m$ are the energy and lifetime of the intermediate state $m$, $M_{fm}$ ($M_{mi}$) is the matrix element of the optical $f\gets m$ ($m\gets i$) transition, and $M_{ep}$ is the matrix element quantifying the exciton-phonon coupling strength.
In our measurements, X$^0$ and X$^-$ are near resonant to the outgoing Raman modes.
Thus, the exciton-Raman detuning (outgoing Raman resonance) dominates the detuning-dependent Raman intensities.
We thereby fit the normalized Raman intensities using
\begin{eqnarray}
	&I'_{p}=\displaystyle\sum_{X}g_{X,p}R_{X,p}, X=\{\mathrm{X}^0, \mathrm{X}^-\},\ p=\{\mathrm{A}_{1g}, \mathrm{2LA}\} \nonumber\\
	&R_{X,p}=1/|\omega_X-(i/2)\gamma_X-\omega_{p}|^2 \nonumber,
\end{eqnarray}
where $g_{X,p}$ reflects the exciton-phonon coupling strength and $R_{X,p}$ is the exciton-Raman detuning modeled by Fermi's golden rule \cite{doi:10.1063/1.4862859,Carvalho2017,gontijo2019}.
The Raman mode energy $\omega_{p}$ are extracted from the Raman spectra.
Since the exciton-Raman resonance deforms the exciton emission line shape \cite{Jones2016,Molas2017}, we extract the exciton energy $\omega_X$ and linewidth $\gamma_X$ from the PL spectra excited by 532 nm-laser (off-resonance, see Sec.~\ref{secs4a} in Supplement \cite{supplement}).

The results of A$_{1g}$ and 2LA measured from the three control cases and two cavities are plotted in Fig.~\ref{f2}(a) as a function of the energy detuning to X$^0$.
Dashed lines represent the resonance arising from the coupling to X$^0$, and dotted lines represent the expected resonance arising from the coupling to X$^-$, $\sim 35\ \mathrm{meV}$ detuned from X$^0$ \cite{Molas2017}.
In three control cases (gray, red, blue), the coupling strength $g_{X,p}$ is nearly constant \cite{doi:10.1063/1.4862859,Carvalho2017,gontijo2019}.
The results reveal that for A$_{1g}$, only $g_{\mathrm{X}^-,\mathrm{A}{1g}}$ has a significant amplitude whilst $g_{\mathrm{X}^0,\mathrm{A}{1g}}$ vanishes.
For 2LA both $g_{\mathrm{X}^0,\mathrm{2LA}}$ and $g_{\mathrm{X}^-,\mathrm{2LA}}$ have finite amplitudes.
Another distinctive feature of X$^-$-Raman resonance is the enhanced ratio of X$^-$/X$^0$ emission intensity \cite{Jones2016,Molas2017} as clearly observed in three control cases presented in Fig.~\ref{f2}(b), where exciton emission around 65 K recorded using the 632 nm-laser (dark, close to X$^-$-Raman resonance) are compared with those using the 532 nm-laser (light, off-resonance).
This is due to the doubly resonant Raman scattering depicted in Fig.~\ref{f2}(c), where X$^0$ is converted to X$^-$ by the phonon scattering \cite{Jones2016,PhysRevLett.122.217401,PhysRevB.102.125410}.
In contrast to the three control cases, Raman spectra recorded from the cavities exhibit entirely different behaviors that are traced to cavity vibrational phonons.
As shown in Fig.~\ref{f2}(a), $g_{\mathrm{X}^0,\mathrm{A}{1g}}$ is clearly nonzero in two cavities, evidenced by the resonance (dashed peak) which is in contrast absent in the three control cases.
Meanwhile, $g_{\mathrm{X}^0,\mathrm{2LA}}$ (dashed peak) is enhanced in cavities whilst the amplitude of $g_{\mathrm{X}^-,\mathrm{2LA}}$ (dotted peak) vanishes.
Therefore, we conclude that X$^0$-phonon coupling strengths are enhanced in cavities while X$^-$-phonon coupling strengths are suppressed.
The suppression of $g_{\mathrm{X}^-,p}$ is also supported by the exciton emission presented in Fig.~\ref{f2}(b).
The X$^-$ emission enhancement \cite{Molas2017} arising from X$^-$-Raman resonance (Fig.~\ref{f2}(c)) is not observed in the cavity.

We note that ingoing Raman resonances have little impact on the measured intensities, since the laser is far detuned from both exciton peaks (X$^0$-laser resonance corresponds to 60 meV in X$^0$-Raman detuning).
Moreover, here A$_{1g}$ and 2LA contain doubly resonant Raman scatterings \cite{Carvalho2017,gontijo2019,Jones2016}, thereby, the ingoing excitation is not limited to the zone center, as depictd schematically in Fig.~\ref{f2}(c)(d).
For the ideal first-order Raman scattering, the ingoing excitation is limited at the zone center, thereby the ingoing resonance between $\omega_{\mathrm{Laser}}$ and $\omega_{\mathrm{X}^0}$ play a role in the variation of the Raman intensity.
In contrast, in our measurements around the exciton-Raman resonance depicted in Fig.~\ref{f2}(c)(d), only the outgoing section is around the zone center, thereby the outgoing resonances $R_{X,p}$ are expected to dominate the spectral dependencies.
Therefore, we fit the detuning-dependent Raman intensities by $R_{X,p}$ as presented in Fig.~\ref{f2}(a).
Nevertheless, at this point we emphasize that the conclusions above can be directly obtained from raw data, without making this approximation and quantitative fittings as discussed in Sec. \ref{secs3a} in Supplement \cite{supplement}.

\begin{figure}
	\includegraphics[width=\linewidth]{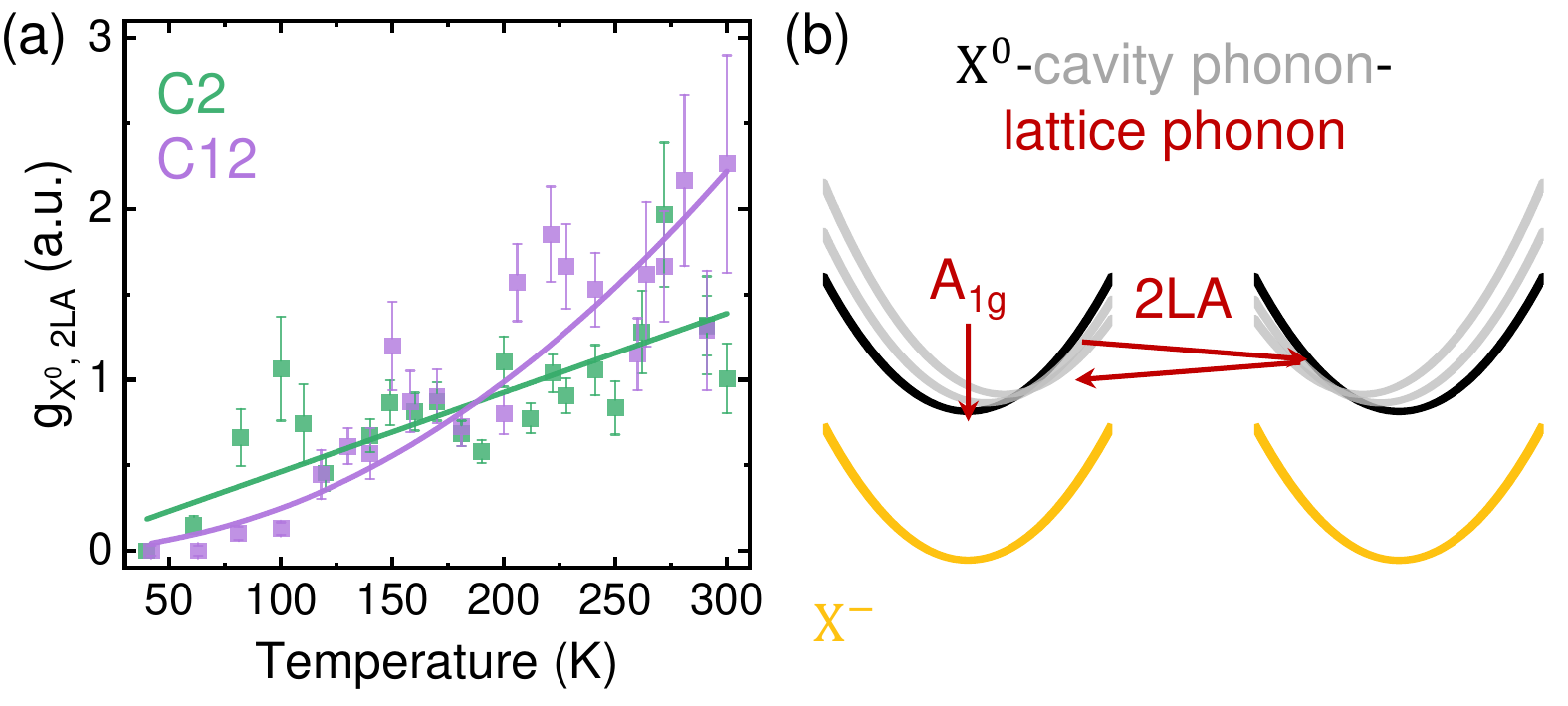}
	\caption{\label{f3}
		(a) $T^N$ dependence of $g_{\mathrm{X}^0,\mathrm{2LA}}$ in cavities.
		(b) Schematic depiction of the exciton-phonon-phonon coupling between X$^0$ (black), cavity phonon (gray replicas) and lattice phonons (dark red arrows).
	}
\end{figure}

Furthermore, the cavity enhanced coupling strengths $g_{\mathrm{X}^0,\mathrm{A}_{1g}}$ and $g_{\mathrm{X}^0,\mathrm{2LA}}$ are found to follow a temperature-dependent power law $T^N$.
This can be seen in Fig.~\ref{f3}(a) that shows $g_{\mathrm{X}^0,\mathrm{2LA}}=I'_{\mathrm{2LA}}/R_{\mathrm{X}^0,\mathrm{2LA}}$ from which we extract $N=1.14\pm0.18$ and $2.21\pm0.23$ for cavity C2 and C12, respectively.
The selective enhancement of $g_{\mathrm{X}^0,p}$ ($p=\{\mathrm{A}_{1g}, \mathrm{2LA}\}$), and the $T^N$ dependence are two key observations in this work, indicating the tripartite coupling between excitons, cavity vibrational phonons and MoS$_2$ lattice phonons.
This exciton-phonon-phonon coupling we envisage is illustrated schematically in Fig.~\ref{f3}(b):  cavity vibrational phonons provide additional intermediate states indicated by the gray replicas that can satisfy the resonance conditions in the scattering of MoS$_2$ lattice phonons (dark red arrows).
The selectivity is due to the vibronic states (gray replicas) that only occur for X$^0$ but not for X$^-$.
We explain this suggest by the weak X$^-$--cavity-phonon coupling originating from the geometry-dependent deformation potentials.
In cavity vibrational modes, the embedded ribbon-shaped MoS$_2$ monolayer extends primarily along the nanobeam axis.
For such uniaxially strained MoS$_2$, Cai et al. \cite{doi:10.1021/ja4109787} calculated the width dependent deformation potential for the electrons ($D_{e}$) and holes ($D_{h}$).
$D_{h}$ was always found to have a magnitude that is approximately twice of $D_{e}$ \cite{doi:10.1021/ja4109787}.
Therefore, the X$^-$--cavity-phonon coupling strength ($\propto 2D_{e}-D_{h}$) is expected to be much smaller than the X$^0$--cavity-phonon coupling strength ($\propto D_{e}-D_{h}$) \cite{2004.14202,PhysRevB.63.155307,PhysRevLett.102.096402}.
This selectivity is further supported by the temperature-dependent linewidth of excitonic PL emission, discussed in SFig.~\ref{sf14} in Supplement \cite{supplement}.

Meanwhile, the $T^N$ dependence is consistent with the Bose occupation of cavity vibrational phonons $n_{cP}={1}/\lbrack e^{{\hbar}\omega_{cP}/\left( k_\mathrm{B} T \right)}-1\rbrack$.
Since the cavity vibrational phonon has low energy (Fig.~\ref{f1}(b)) $\hbar\omega_{cP}\ll k_\mathrm{B} T$, the temperature dependent factor of either Stokes ($n_{cP}+1$) or anti-Stokes ($n_{cP}$) processes is $\approx  k_\mathrm{B} T/\left(\hbar\omega_{cP}\right) \propto T$.
Thereby, $N$ is determined by the number of cavity phonons participating in the tripartite coupling.
Additional evidence for the tripartite coupling is observed from the spatial dependence of Raman enhancement and the temperature dependence of Raman linewidth (anharmonicity) \cite{LIU2019451} as discussed in Sec. III B in Supplement.
Similar vibronic sublevel mediated processes have been reported in photonic \cite{doi:10.1063/1.3671458,Rosser2020,PhysRevLett.126.227401,PhysRevLett.128.087401} and plasmonic systems \cite{PhysRevA.100.043422,B505343A}.
In contrast, other factors do not explain the experimental results.
For example, the control experiments in the supported and suspended case show that both two types of static strain without cavity vibrational modes do not result in the selective Raman enhancement.
Indeed, the static strain is not proportional to temperature, and thereby, it cannot explain the $T^N$ dependence.
The resonant cavity optical mode could of course enhance the Raman intensity by increasing the local optical field for the laser \cite{C2AN35722D,doi:10.1080/05704928.2019.1661850} or exciton spontaneous emissions \cite{doi:10.1080/05704928.2019.1661850,PhysRevA.100.043422,doi:10.1021/acs.jpcc.0c11421}.
However, these effects are strongly dependent on the detuning of the cavity optical mode to the laser or excitonic transitions.
In our experiments, the excitation laser and Raman modes are far detuned from the cavity optical mode ($>20\ \mathrm{meV}$), and the detuning is nearly temperature independent.
The coupling between X$^0$ and the cavity optical mode is negligible \cite{PhysRevLett.128.237403}, and no dependence on their detuning is observed in Fig.~\ref{f3}(a).
Therefore, several pieces of evidence all indicate that the cavity optical mode does not play the major role in determining the observed Raman enhancement.

In summary, we performed spatially resolved Raman spectroscopy to demonstrate how excitons, MoS$_2$ lattice phonons and cavity vibrational phonons couple to govern the exciton-phonon scattering in 2D-material nanophotonic cavities.
The selectivity to the neutral exciton indicates such phononic technology can be applied to control light-matter interactions based on different electronic transitions (excitonic sorting).
Moreover, our results are obtained for thermally excited vibrational modes without external driving \cite{doi:10.1021/acs.nanolett.0c05089,doi:10.1021/nl501413t}.
Therefore, our results indicate that the phononic hybridization between lattice and nanomechanical modes has an intrinsic strong impact on excitonic photophysics and light-matter interactions in 2D-material nanophotonic systems.

All authors gratefully acknowledge the German Science Foundation (DFG) for financial support via grants FI 947/8-1, DI 2013/5-1 and SPP-2244, as well as the clusters of excellence MCQST (EXS-2111) and e-conversion (EXS-2089). C. Q. and V. V. gratefully acknowledge the Alexander v. Humboldt foundation for financial support in the framework of their fellowship programme.

C. Q. and V. V. contributed equally to this work.


%

\clearpage

\section*{Supplementary Information}
\setcounter{figure}{0}
\renewcommand{\figurename}{SFig.}


\section{\label{secs1} Setup and Methods}

\subsection{\label{secs1a}Sample Fabrication}

\begin{figure}
	\includegraphics[width=\linewidth]{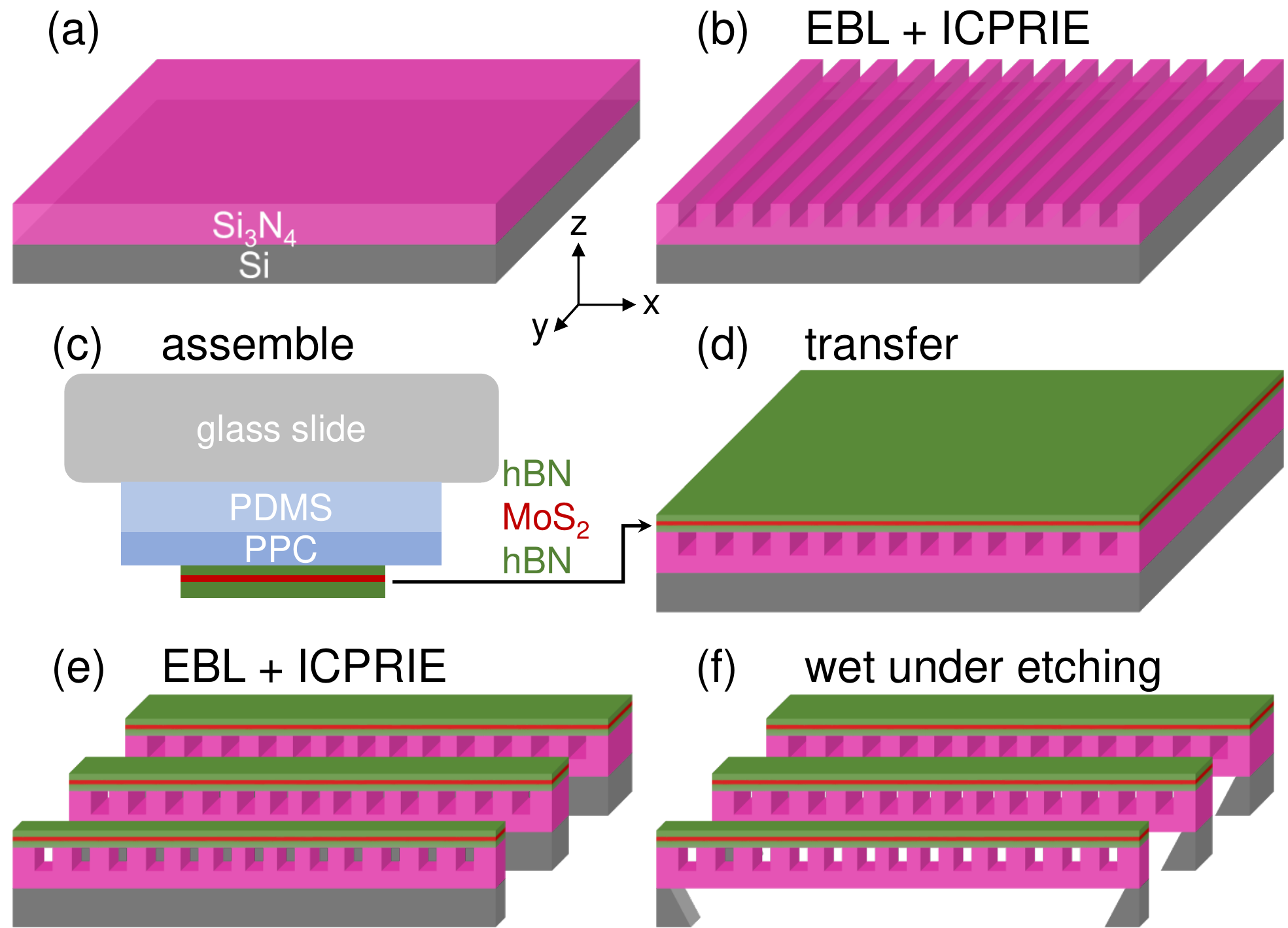}
	\caption{\label{sf1}
		Design and fabrication of the sample \cite{PhysRevLett.128.237403}.
		(a) Si$_3$N$_4$/Si substrate.
		(b) EBL and ICPRIE to fabricate periodic nanoscale trenches.
		(c) Viscoelastic method to assemble the hBN/MoS$_2$/hBN heterostructure.
		(d) Transfer the heterostructure on top of the pre-etched Si$_3$N$_4$.
		(e) Second EBL and ICPRIE to divide the nanobeams.
		(f) Wet etching to remove the bottom Si.
	}
\end{figure}

The fabrication processes are schematically depicted in SFig.~\ref{sf1}.
Firstly, as shown in SFig.~\ref{sf1}(a) we prepare and clean the Si$_3$N$_4$/Si substrate which is from Active Business Company GmbH.
The Si$_3$N$_4$ is grown by low pressure chemical vapor deposition (LPCVD) and has the thickness of 200 nm.
Then we use e-beam lithography (EBL) and inductively coupled plasma reactive ion etching (ICPRIE) to etch the periodic nanoscale trenches in Si$_3$N$_4$.
The EBL machine is eLINE from Raith GmbH, and the ebeam resist is AR-P 6200 from Allresist GmbH.
We use e-beam at 30 kV with the dose $\mathrm{150\ \mu C/cm^2}$ to pattern the resist with the thickness of 270 nm in this step.
The ICPRIE machine is PlasmaPro 80 from Oxford Instruments.
We use SF$_6$ and C$_4$F$_8$ with ratio 3:2, pressure 13.5 mTorr, HF power 15 W and ICP power 220 W for ICPRIE.
After the first nanofabrication, we prepare and transfer the hBN/MoS$_2$/hBN heterostructure on top of the pre-etched Si$_3$N$_4$ as shown in SFig.~\ref{sf1}(c)(d).
The bulk hBN and MoS$_2$ we use for exfoliation are from HQ Graphene.
We use the PVA-assisted method to exfoliate huge hBN and MoS$_2$ flakes \cite{Huang2020} and use PPC stamp for the viscoelastic dry transfer \cite{Pizzocchero2016}.
The stacking and transfer is implemented in a home-built machine.
Finally, we use the second EBL and ICPRIE to divide the nanobeams as shown in SFig.~\ref{sf1}(e) following by the wet etching to remove the bottom Si as shown in SFig.~\ref{sf1}(f).
In the second EBL, we use the same voltage but the dose $\mathrm{125\ \mu C/cm^2}$ for the resist with the thickness of 480 nm.
The gases and parameters in the second ICPRIE are same to the first in SFig.~\ref{sf1}(b).
We use 25$\%$ TMAH solution for the wet under etching in SFig.~\ref{sf1}(f).

\subsection{\label{secs1b}Raman Measurement}

Raman spectra in this work are measured by a confocal micro-Raman system.
The excitation cw-laser has the wavelength of 632 nm.
The laser is focused by the objective with a magnification of 100 and a NA of 0.75 into a spot size $\sim1\ \mathrm{\mu m^2}$.
The temperature of sample is controlled by the liquid helium flow and heater.
The sample position is controlled by a three-dimensional xyz nanopositioner.
The signal is collected by a matrix array Si CCD detector in the spectrometer with a focal length of 0.55 m and a grating of 1200 grooves per mm.

\begin{figure}
	\includegraphics[width=\linewidth]{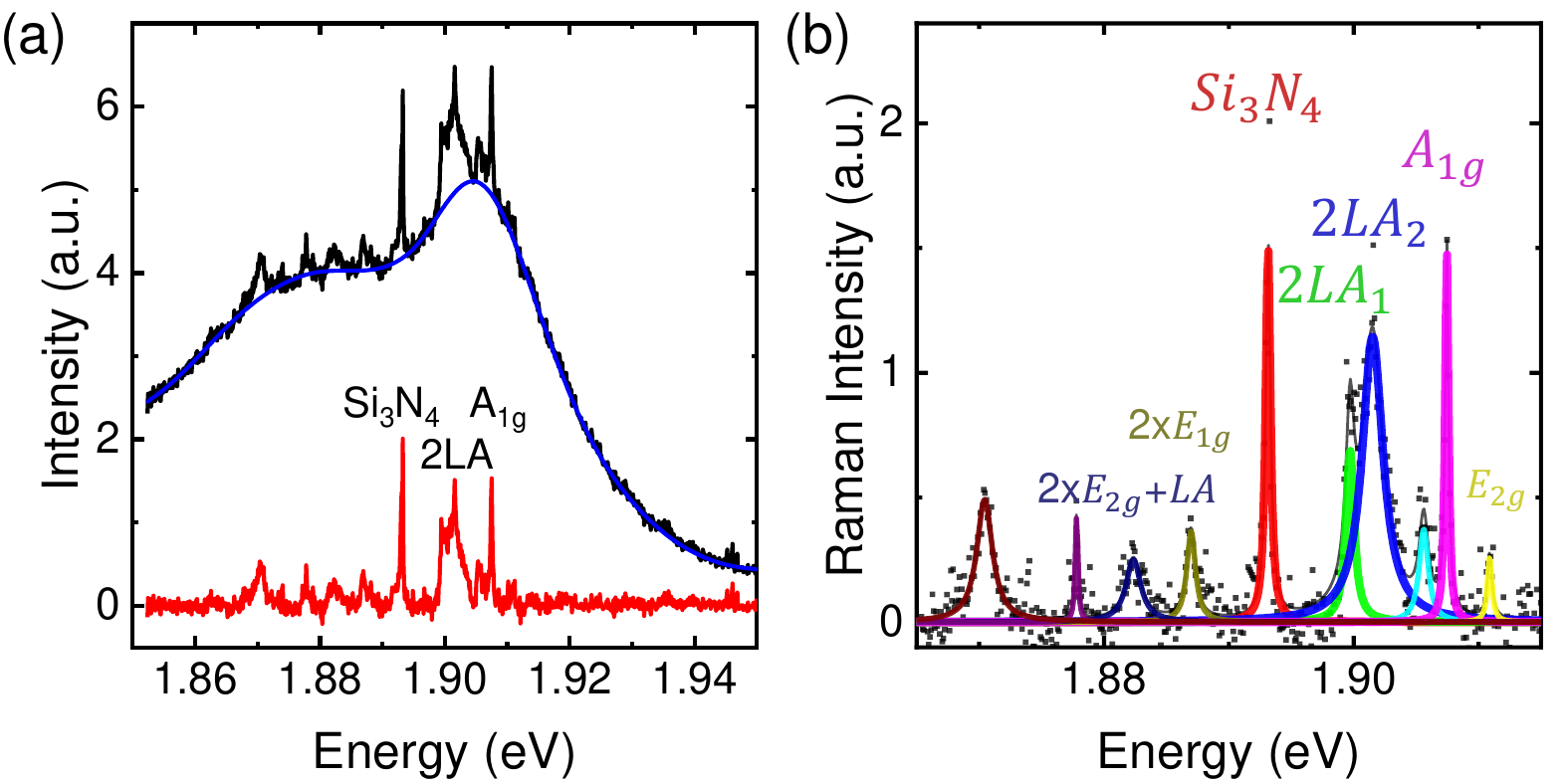}
	\caption{\label{sf2}
		Subtraction and fitting of Raman spectra.
		(a) One example of the raw data (black) and the corresponding Raman spectrum (red) after subtracting the emission baseline (blue).
		(b) Multi-Lorentz fitting. The intensity of 2LA discussed in this work is the sum of two peaks (blue and green).
	}
\end{figure}

As shown in Fig.~1(c), the resonant Raman spectroscopy produce both exciton emissions and Raman signals superimposed in spectra.
We present the methods for the subtraction and fitting in SFig.~\ref{sf2}.
The black line in SFig.~\ref{sf2}(a) is the raw data collected by the spectrometer.
The minor narrow peaks are from Raman signals, and the broad peaks are from the emission of X$^0$ and X$^-$.
We sketch the emission baseline based on the feet of minor (Raman) peaks as the blue line.
Then the Raman spectra are extracted by subtracting the emission baseline from the raw data, denoted by the red line in SFig.~\ref{sf2}(a).
Such subtraction method is widely applied in resonant Raman spectroscopy \cite{Molas2017}.
We then fit the Raman spectra by multi Lorentz peaks as shown in SFig.~\ref{sf2}(b).
Besides A$_{1g}$ and 2LA, we also observe other Raman peaks.
The known Raman peaks are labelled in SFig.~\ref{sf2}(b).
However, other peaks besides A$_{1g}$ and 2LA are too weak to distinguish when they are strongly detuned to X$^0$ and X$^-$.
Therefore, we mainly focus on A$_{1g}$ and 2LA in this work.

\subsection{\label{secs1c}PL Measurement}

\begin{figure}
	\includegraphics[width=\linewidth]{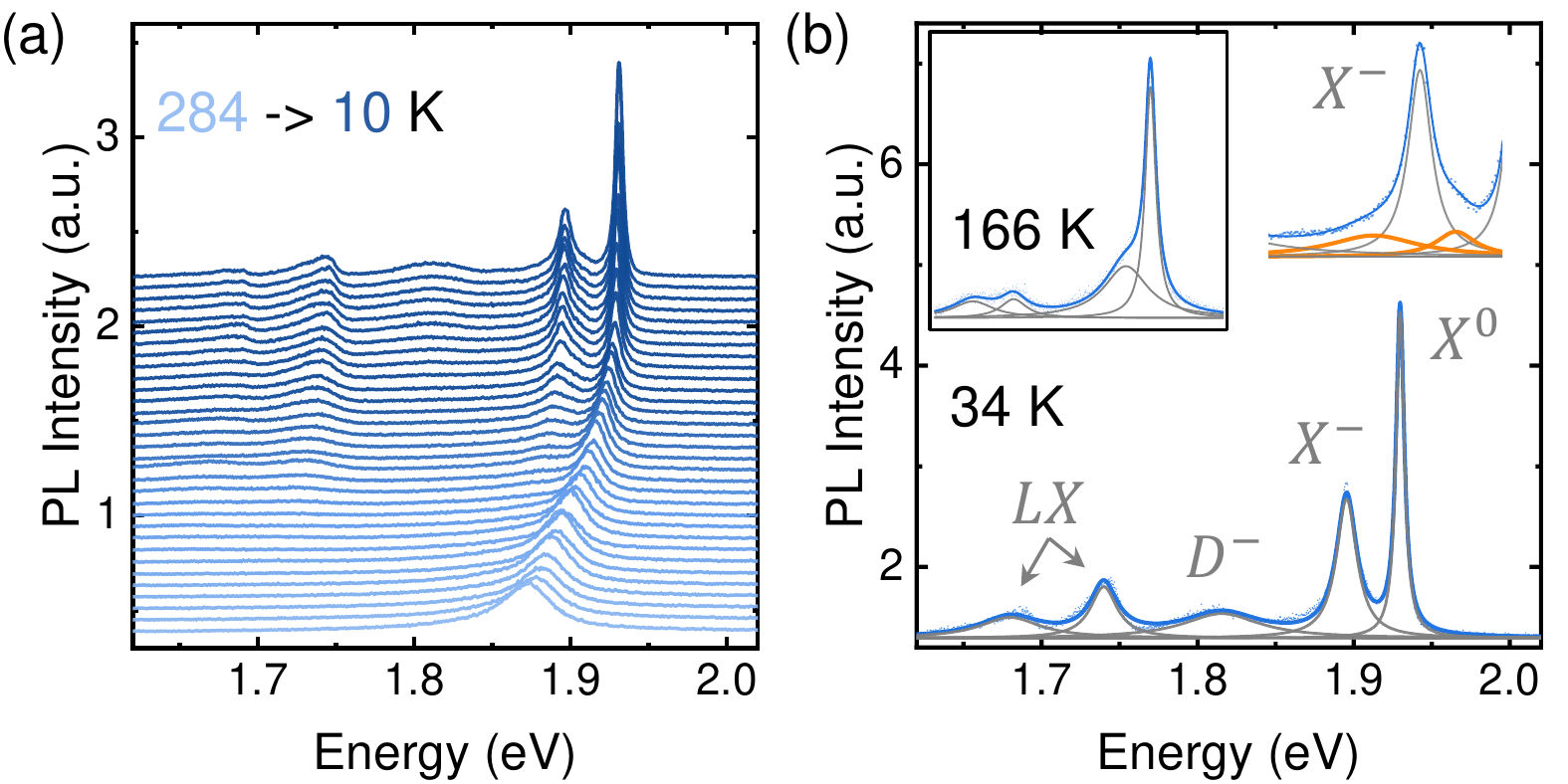}
	\caption{\label{sf3}
		(a) Temperature-dependent PL spectra measured in the suspended case.
		(b) Multi-Lorentz fittings of the PL spectra at 34 and 166 K as examples.
		PL peaks from neutral exciton X$^0$, intravalley trion X$^-$, dark trion D$^-$ and localized excitons LX (two peaks) are observed \cite{He2020,PhysRevMaterials.3.051001}.
	}
\end{figure}

We also study the exciton-phonon coupling using PL spectroscopy.
The confocal micro-PL setup is generally same to the micro-Raman setup, but with the excitation cw-laser at 532 nm wavelength.
The excitation laser has the spot size $\sim 1\ \mathrm{\mu m}$ and the power $\sim 30\ \mathrm{\mu W}$.
In this case, the Raman modes are strongly detuned from the exciton emissions.
The temperature $T$ is similarly controlled by the liquid helium flow and heater.
We collect the exciton emission spectra by the spectrometer.
Typical temperature-dependent PL spectra measured from the suspended case are presented in SFig.~\ref{sf3}(a).
We use multi Lorentz fittings to extract the properties of exciton emission peaks including neutral exciton X$^0$, intravalley trion X$^-$, dark trion D$^-$ and localized excitons LX (two peaks) \cite{He2020,PhysRevMaterials.3.051001} as presented in SFig.~\ref{sf3}(b).

However, at high temperature $T>150\ \mathrm{K}$, the X$^-$ and D$^-$ peak merges such as shown by the spectra at 166 K in SFig.~\ref{sf3}(b) inset.
As a result, the fitting accuracy of X$^-$ linewidth $\gamma_{X^-}$ at high $T$ is limited.
Moreover, trion emission in monolayer MoS$_2$ has fine structures.
Besides the major emission peak from intravalley singlet trion X$^-$, there are also minor peaks from intervalley singlet trion, intervalley triplet trion and many-body states \cite{PhysRevB.105.L041302}.
As an alternative fitting method, these minor trion peaks can be extracted at low $T$, such as shown by the orange peaks in SFig.~\ref{sf3}(b).
In this work, we extract $\gamma_{X^-}$ by the fitting without the minor trion peaks.
Differences between the two fitting methods (with or without minor trion peaks) are added to the uncertainties of $\gamma_{X^-}$.

\section{\label{secs2} Calculation of Cavity Modes}

\begin{figure}
	\includegraphics[width=\linewidth]{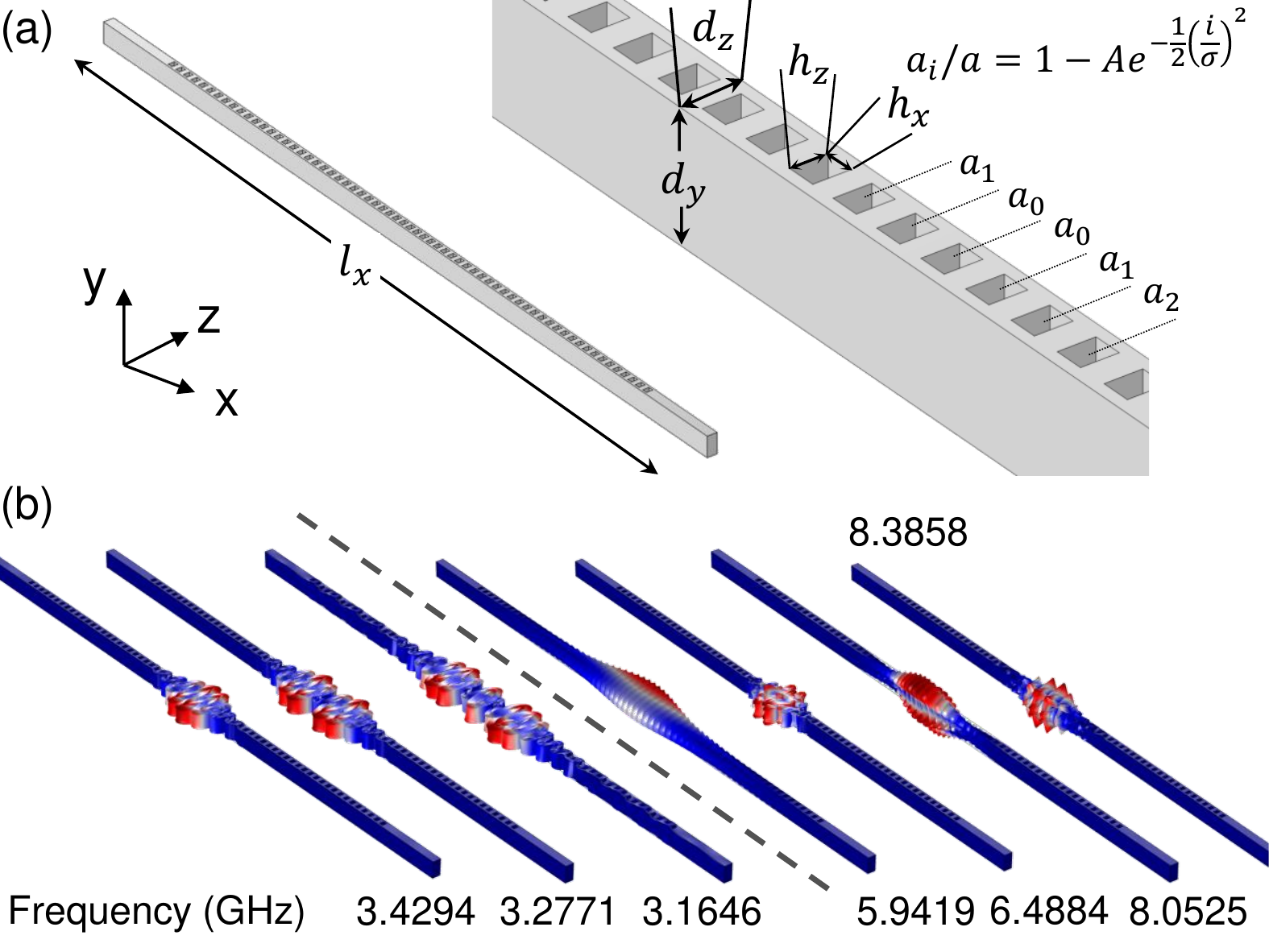}
	\caption{\label{sf4}
		(a) Cavity structure \cite{PhysRevLett.128.237403}.
		(b) Typical confined vibrational modes calculated by FEM method.
		The first three modes are already plotted in Fig.~1(b).
	}
\end{figure}

3D finite element method (FEM) method is used to calculate the phononic modes of cavity.
Detailed parameters of cavity structure are denoted in SFig.~\ref{sf4}(a).
The whole nanobeam has the length $l_x=20\ \mathrm{\mu m}$ and depth $d_z=250\ \mathrm{nm}$.
The width is $d_y=500\ \mathrm{nm}$ in the calculation but varies between $d_y=300-530\ \mathrm{nm}$ in experiments.
Nanoscale trenches have length $h_x=120\ \mathrm{nm}$ and depth $h_z=150\ \mathrm{nm}$, and follow a Gaussian distribution in spatial with the separation between trenches $a_i/a=1-A\cdot \mathrm{exp}(-i^2/(2\sigma^2))$.
$a=250\ \mathrm{nm}$ is the lattice constant and $A=0.1,\sigma=4$ define the smoothly varying photonic and phononic confinement.
The photonic modes have been reported previously \cite{PhysRevLett.128.237403}.
The calculation results of confined phononic (vibrational) modes are presented in SFig.~\ref{sf4}(b).
As shown, the frequency of confined vibrational modes is $\sim \mathrm{GHz}$.
The mode frequency and distribution are consistent to similar structures reported in previous works \cite{10.1038/nature08061,10.1038/nature08524,RevModPhys.86.1391}.
Since the phonon energy $\hbar\omega_{cP}\ll k_B T$, the Bose factor of the cavity vibrational phonons $n_{cP}={1}/\lbrack e^{{\hbar}\omega_{cP}/\left( k_B T \right)}-1\rbrack\approx  k_B T/\left(\hbar\omega_{cP}\right)$ is proportional to the temperature $T$.
We note that since the optical and mechanical properties of hBN are not well studied yet and strongly depend on the material quality such as the defect concentration \cite{Thomas_2015}, we set all material to Si$_3$N$_4$ in the calculation.
This might introduce inaccuracy in the calculated eigenfrequencies, but the order of magnitudes $\sim \mathrm{GHz}$ should be correct.
Even the actual frequency of phononic modes is two order of magnitude larger e.g. 300 GHz, in our measurement range $T>50\ \mathrm{K}$ we have $\hbar\omega_{cP}/\left(k_B T\right)<0.29$ thus the temperature dependence of the Bose factor is still approximately linear.

\begin{figure}
	\includegraphics[width=\linewidth]{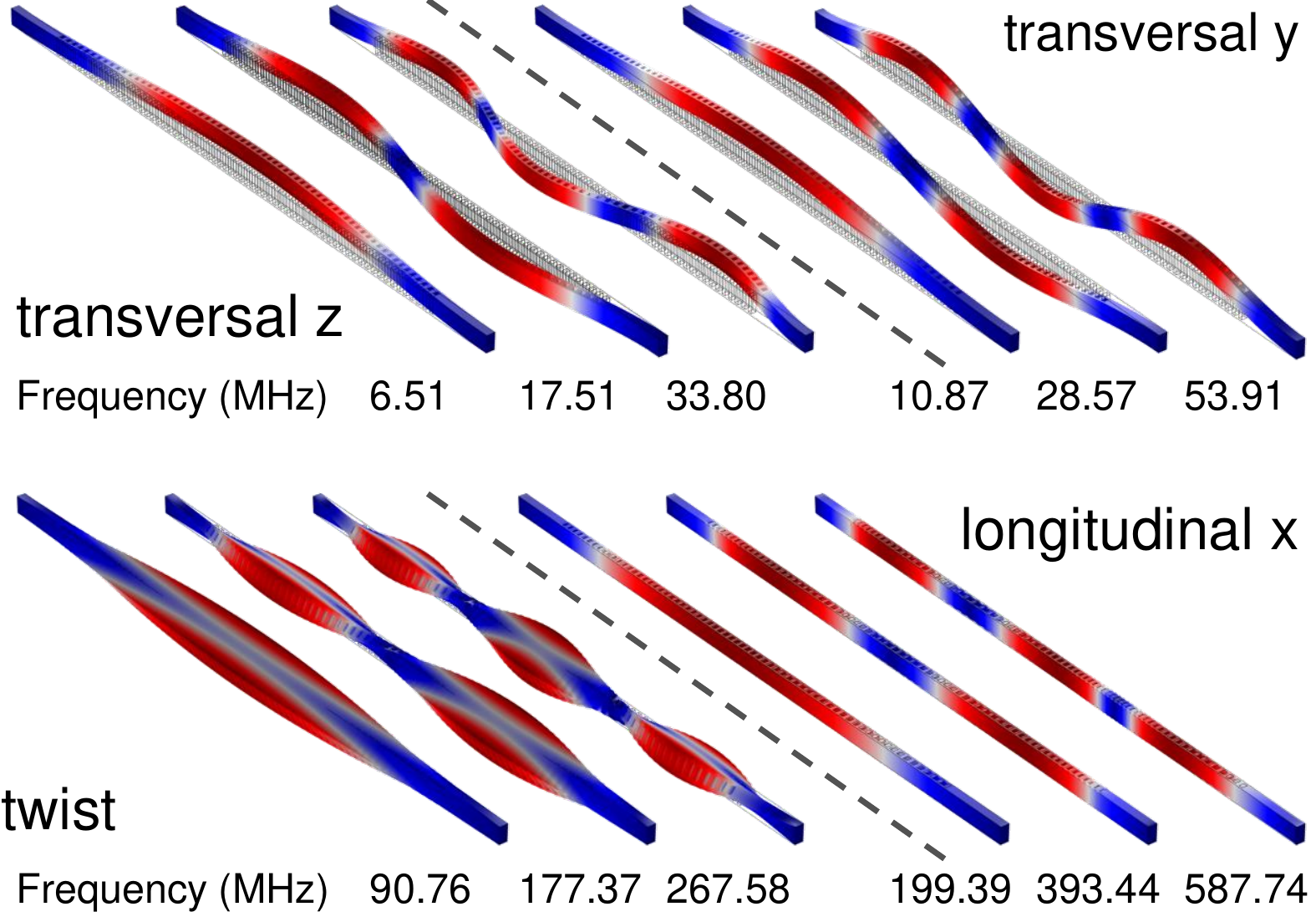}
	\caption{\label{sf5}
		Typical beam vibrational modes.
	}
\end{figure}

\begin{figure}
	\includegraphics[width=\linewidth]{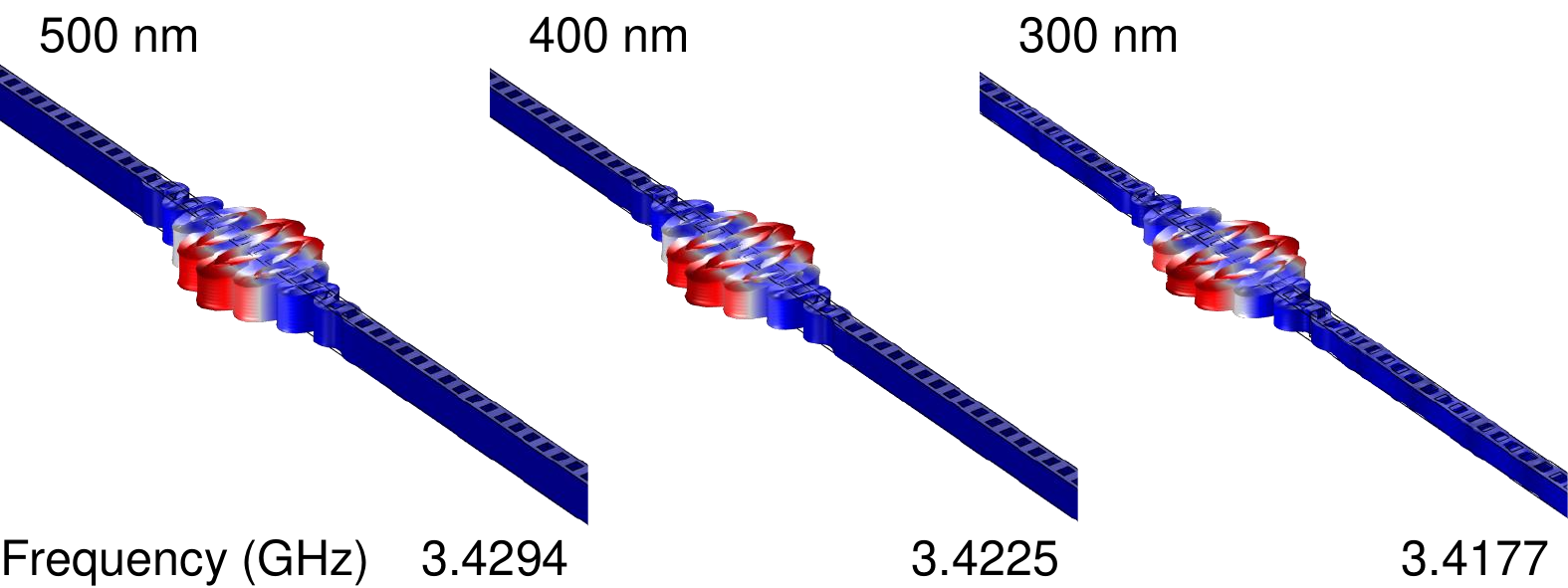}
	\caption{\label{sf6}
		Calculated vibrational modes for cavities with the nanobeam width $d_y$ of 500, 400 and 300 nm.
		They have the similar profile and slightly different frequencies.
	}
\end{figure}

In addition, the nanobeam cavity also supports the vibrational modes of the whole nanobeam as presented in SFig.~\ref{sf5}.
The beam vibrational modes have the frequency $\sim \mathrm{MHz}$, even lower than the confined vibrational modes.
Meanwhile, the nanobeam width $d_y$ rarely affects the cavity mode profile and slightly modifies the mode frequency, as presented in SFig.~\ref{sf6}.
The key point is that the Bose factor of all these cavity modes is $\propto T$.
This linear dependence provides the explanation for the $T^N$ dependence in the Raman enhancement.

\section{\label{secs3}Additional Raman Results}

\subsection{\label{secs3a}Alternative Analysis Methods}

\begin{figure}
	\includegraphics[width=\linewidth]{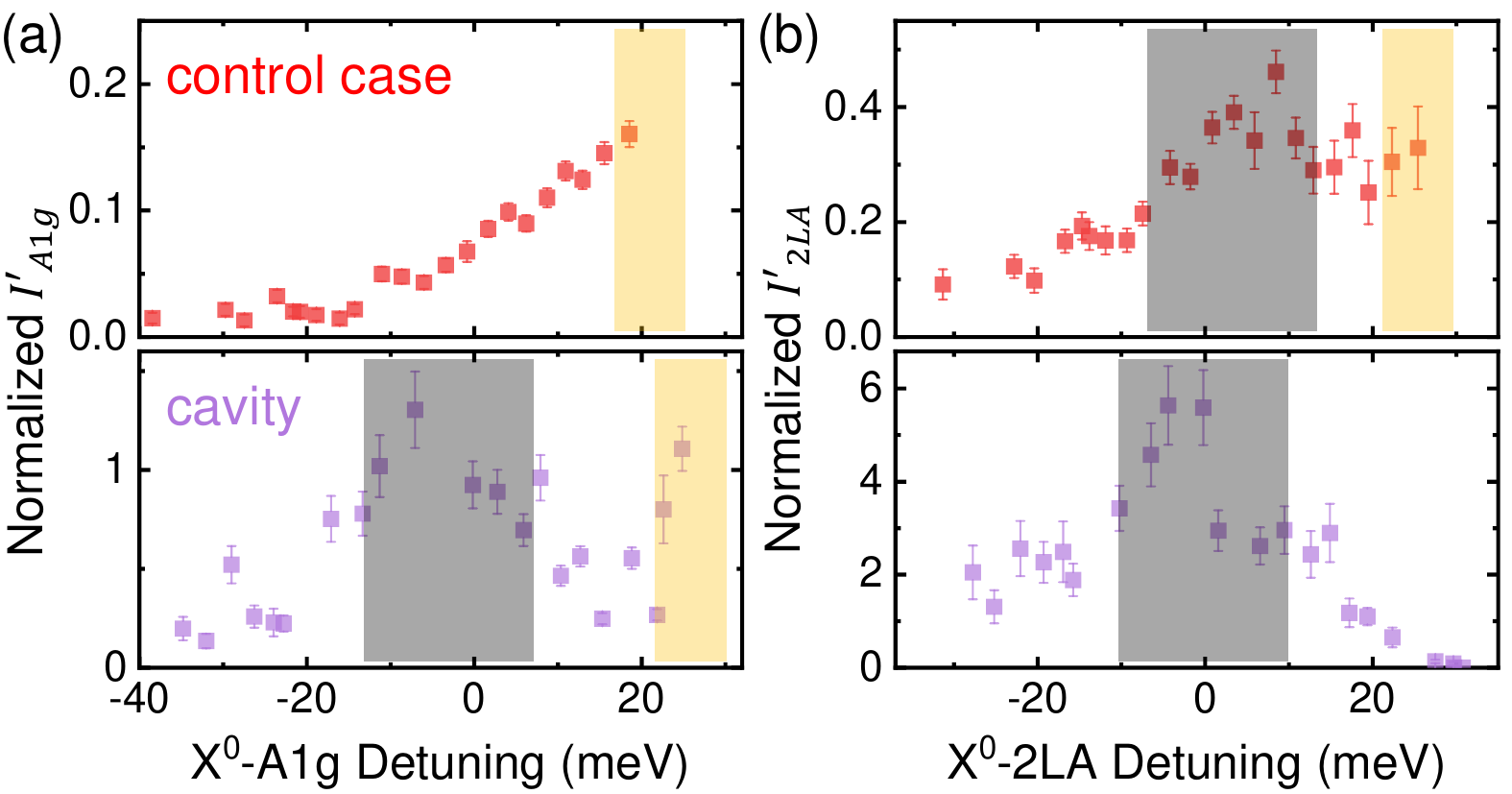}
	\caption{\label{sf7}
		Analysis of raw Raman datasets without quantitative fittings.
		The maximum in the gray region is surely from X$^0$-phonon resonance.
		The maximum in the yellow region surely contains contribution from X$^-$-phonon resonance.
	}
\end{figure}

We emphasize that the selective Raman enhancement does not rely on the quantitative fitting method and can be directly obtained from the raw data.
Generally, we obtain two maximums in the Raman intensities, denoted by the gray and yellow regions in SFig.~\ref{sf7}.
The maximum in the gray region around zero detuning is surely from the X$^0$-Raman resonance.
Comparing the cavity to the control cases, the enhancement of this maximum is clearly observed.
Therefore, the first conclusion of enhanced X$^0$-induced Raman scattering in cavities, is obtained.
One might wonder about the maximum in the yellow region.
This maximum in SFig.~\ref{sf7}(b) surely contains the contribution from the X$^-$-2LA detuning, but might also be contributed by the X$^0$-LA or X$^0$-Laser detuning.
Nonetheless, we do not need to care about how much contribution is from which detuning, because this maximum of 2LA in yellow region completely vanishes in cavities.
This means all contributions vanish.
Therefore, the second conclusion of suppressed X$^-$-induced Raman scattering in cavities, is also obtained.

\begin{figure}
	\includegraphics[width=\linewidth]{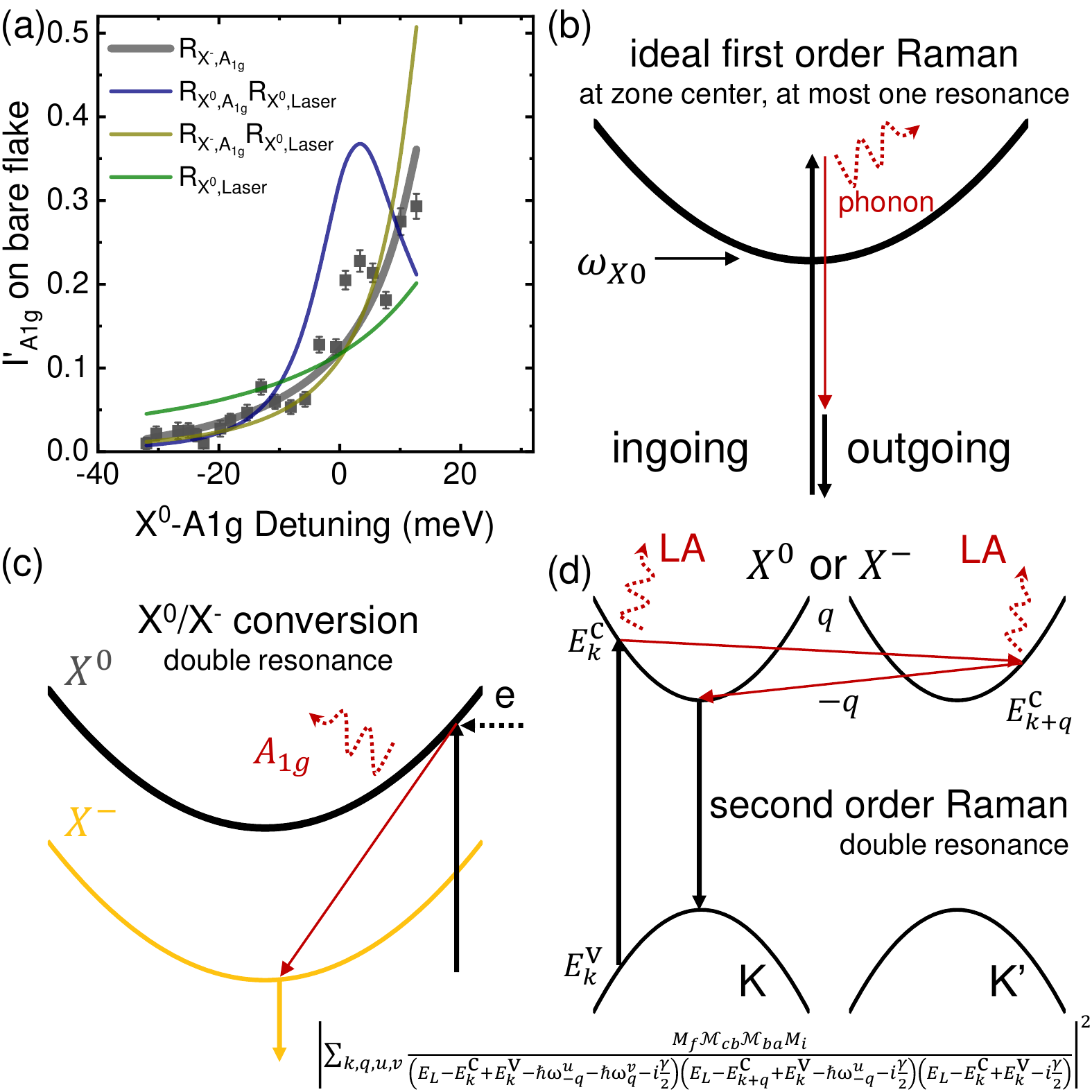}
	\caption{\label{sf8}
	Schematic of Raman scatterings.
	(a) $I'_{\mathrm{A}{1g}}$ from the bare flake case.
	By comparison $R_{X^-,\mathrm{A}{1g}}$ (gray line) is the best fitting.
	(b) The ideal first-order Raman scattering limited at zone center.
	(c) The phonon-assisted X$^0$/X$^-$ conversion \cite{Jones2016}.
	(d) The second-order 2LA scattering \cite{Carvalho2017}.
	}
\end{figure}

In Fig.~2(a) we use the exciton-Raman detuning $R_{X,p}$ to fit the Raman intensity since this detuning dominates the intensity variation in our measurements.
Hereby, we plot the data of A$_{1g}$ in the bare flake case in SFig.~\ref{sf8}(a) and discuss in detail.
The gray line denotes the fitting by $R_{X^-,\mathrm{A}{1g}}$ as used in the main paper.
The dark blue line in SFig.~\ref{sf8}(a) is the prediction by $R_{X^0,\mathrm{A}{1g}}R_{X^0,Laser}$ which corresponds to the ideal first-order Raman scattering schematically depicted in SFig.~\ref{sf8}(b).
In this case, both the ingoing and outgoing section are limited at the zone center, and thereby, the Raman intensity is described by the multiple of the ingoing detuning $R_{X^0,Laser}$ and the outgoing detuning $R_{X^0,\mathrm{A}{1g}}$.
Obviously, SFig.~\ref{sf8}(b) is not our case.
Even if only considering $R_{X^0,Laser}$ (dark green line in SFig.~\ref{sf8}(a)), the variation is much smaller than the experimental data due to the large X$^0$-Laser detuning.
Moreover, only considering $R_{X^0,Laser}$ is contradictory, since  $R_{X^0,\mathrm{A}{1g}}$ and $R_{X^0,Laser}$ are from one scattering process (SFig.~\ref{sf8}(b)), thereby one term exists while the other term vanishes is contradictory.

The intensity of A$_{1g}$ contains signals from the doubly resonant Raman scattering \cite{Jones2016} depicted SFig.~\ref{sf8}(c).
In the doubly resonant scattering, the outgoing section (downward yellow arrow) is around the zone center thus follows $R_{X^-,\mathrm{A}{1g}}$.
In contrast, in the ingoing section (upward black arrow), the X$^0$ created by the laser is not at the zone center, thus does not follow $R_{X^0,Laser}$.
Indeed, the ingoing section creates X$^0$ with the laser energy $\omega_{Laser}$ at a real energy level.
Therefore, we think this probability is dominated by $1+n_{Laser}$, since excitons in the valley follows the Bose distribution with the Bose factor $n_{Laser}={1}/\lbrack e^{{\hbar}\left(\omega_{Laser}-\omega_{X0}\right)/\left( k_B T \right)}-1\rbrack$ \cite{PhysRevB.98.020301}.
In our measurements, $1+n_{Laser}\approx 1$ since $\hbar\omega_{Laser}-\hbar\omega_{X0}\gg k_B T$.
Therefore, the ingoing section has little impact on the Raman intensity.
As shown in SFig.~\ref{sf8}(a), $R_{X^-,\mathrm{A}{1g}}$ fits the experimental data better than $R_{X^-,\mathrm{A}{1g}}R_{X^0,Laser}$ (dark yellow line).

Similarly, the intensity of 2LA contains signals from the doubly resonant Raman scattering \cite{Carvalho2017} depicted in SFig.~\ref{sf8}(d).
The ingoing section is also not limited at zone center.
Strictly, the Raman intensity is very complex and consists of scattering with all possible intermediate states, such as the sum equation in SFig.~\ref{sf8}(d) reproduced from \cite{Carvalho2017}.
However, in our measurements the outgoing section is around resonance, thus $R_{X,p}$ plays the major role in the Raman intensity variation and is used to fit the detuning dependence.
The validity of our fitting method is also supported by the comparison to the exciton-phonon coupling strength extracted from PL data as discussed later in Sec.~\ref{secs4a}.

\begin{figure}
	\includegraphics[width=\linewidth]{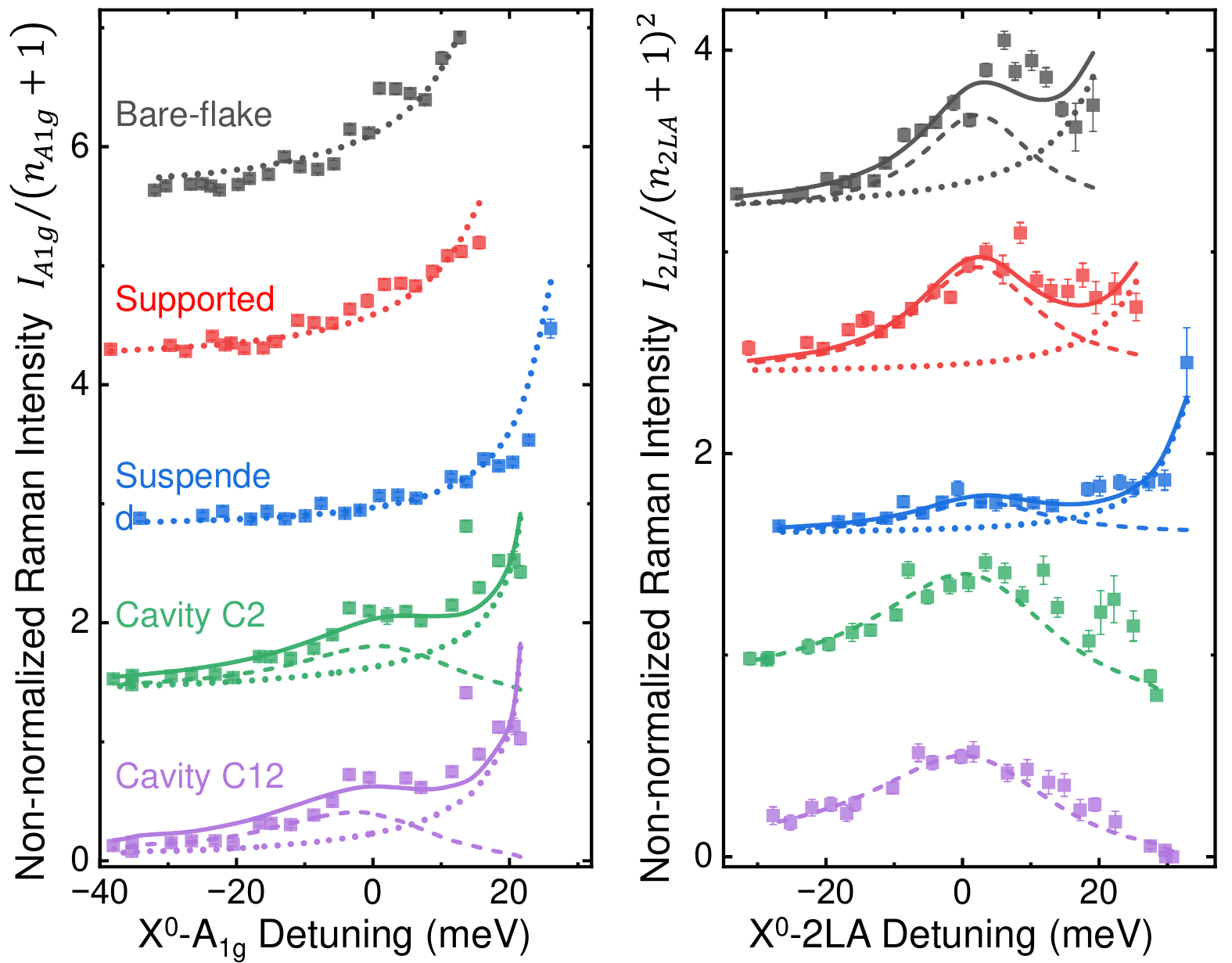}
	\caption{\label{sf9}
		Raman intensities $I_{\mathrm{A}_{1g}}/(n_{\mathrm{A}_{1g}}+1)$ and $I_{\mathrm{2LA}}/(n_{\mathrm{LA}}+1)^2$ without the normalization to Si$_3$N$_4$ peak.
		Conclusions are same to Fig.~2 in the main paper.
	}
\end{figure}

\begin{figure*}
	\includegraphics[width=0.8\linewidth]{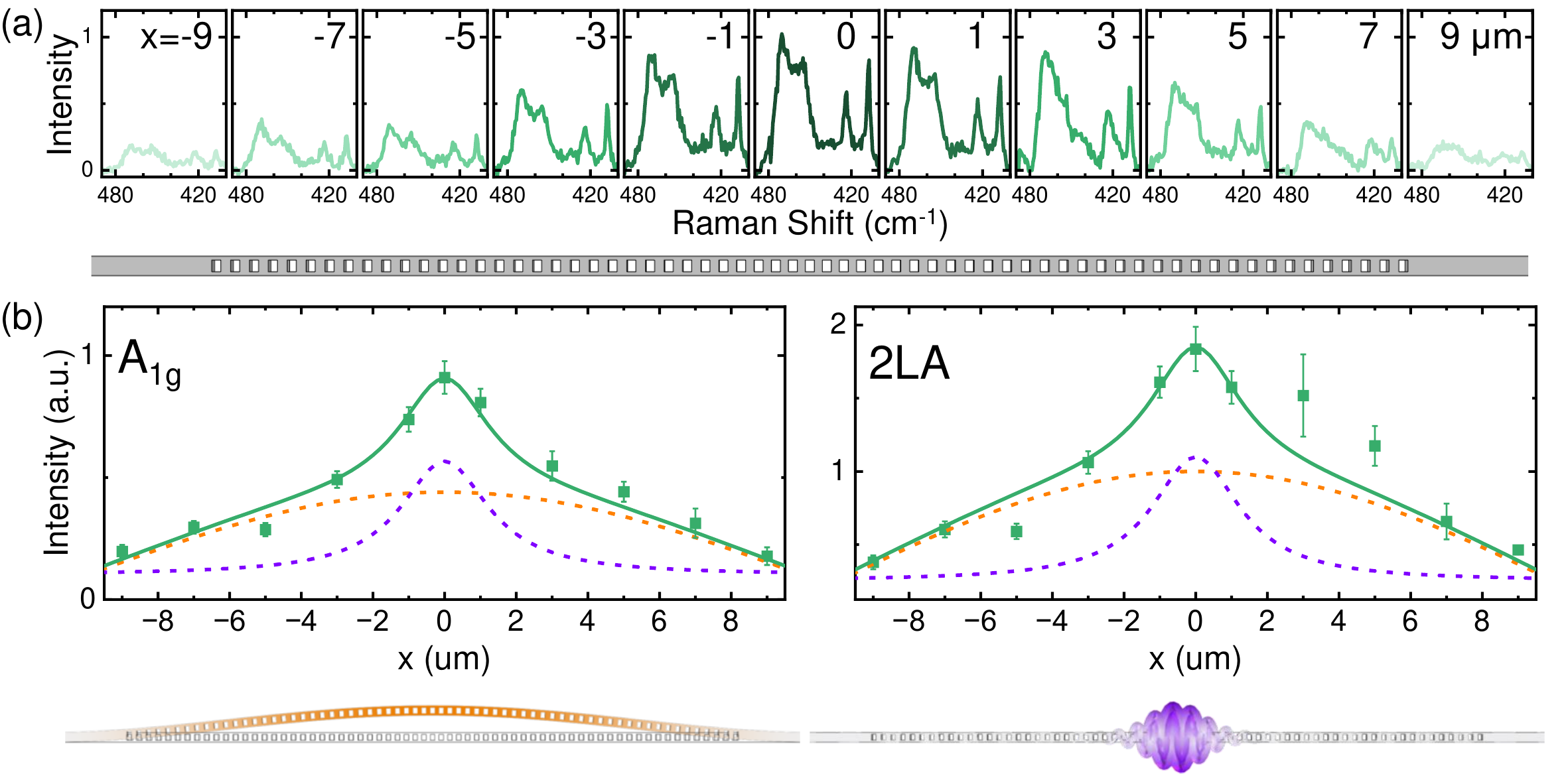}
	\caption{\label{sf10}
		Position dependence of the Raman enhancement.
		(a) Position-dependent Raman spectra along cavity C2, measured at $T=268\ \mathrm{K}$.
		(b) Extracted Raman intensities of A$_{1g}$ and 2LA.
		The position dependence indicates that both beam (orange) and confined (violet) vibrational modes contribute to the Raman enhancement.
	}
\end{figure*}

We normalized the intensity of A$_{1g}$ and 2LA by the Si$_3$N$_4$ peak to remove potential noises in the measurements \cite{C9NR02447F}.
In addition, the normalization provides the comparison of Raman intensities between different cases.
It makes no sense to compare the absolute Raman intensity between difference cases, since the laser spot size $1\ \mathrm{\mu m}$ is much larger than the nanobeam cavity width $d_y\leq 0.5\ \mathrm{\mu m}$ and $d_y$ varies between different cavities.
This means the size of MoS$_2$ excited by the laser varies between different cases.
Nevertheless, we emphasize that our conclusions do not rely on the normalization.
We present the Raman intensities of A$_{1g}$ and 2LA without the normalization to Si$_3$N$_4$ peak in SFig.~\ref{sf9}.
As shown, the two key results are same to the normalized intensities in Fig.~2(a), including the selective enhancement of X$^0$-induced scattering and the $T^N$ dependence of the enhancement in cavities.

\subsection{\label{secs3b}Additional Evidence for the Coupling}

Here we present the position dependence of the Raman spectra along the nanobeam in SFig.~\ref{sf10}(a).
The Raman enhancement is generally centralized.
The extracted intensity of A$_{1g}$ and 2LA are presented in SFig.~\ref{sf10}(b), indicating that both beam (orange) and confined (violet) vibrational modes contribute to the Raman enhancement.
The beam vibrational modes have broader spatial distribution thus introduce an enhancement broad in spatial.
In contrast, the confined vibrational modes introduce an enhancement narrow in spatial.
The sum of two are denoted by the fitting lines in SFig.~\ref{sf10}(b).
We note that the fitting is based on the typical modes and qualitative, since there exist multiple modes for both beam (SFig.~\ref{sf5}) and confined (SFig.~\ref{sf4}) vibrations.

The confined vibrational modes have smaller mode volume but also smaller population (high frequency $\sim$GHz).
In contrast, the beam vibrational modes have the larger population (low frequency $\sim$MHz) but larger mode volume.
The exciton-phonon coupling will increase with the phonon population but decreases with the mode volume \cite{Chen2015}.
Thus, it is a reasonable result that both two types of vibrational modes contribute.
We emphasize that the beam and confined vibrations are both cavity vibrational phonons.
The results in SFig.~\ref{sf10} reveal that the tripartite coupling can occur with different kinds of nanomechanical modes, therefore, the phononic hybridization can be applied to a wide range of nanosystems.

\begin{figure}
	\includegraphics[width=\linewidth]{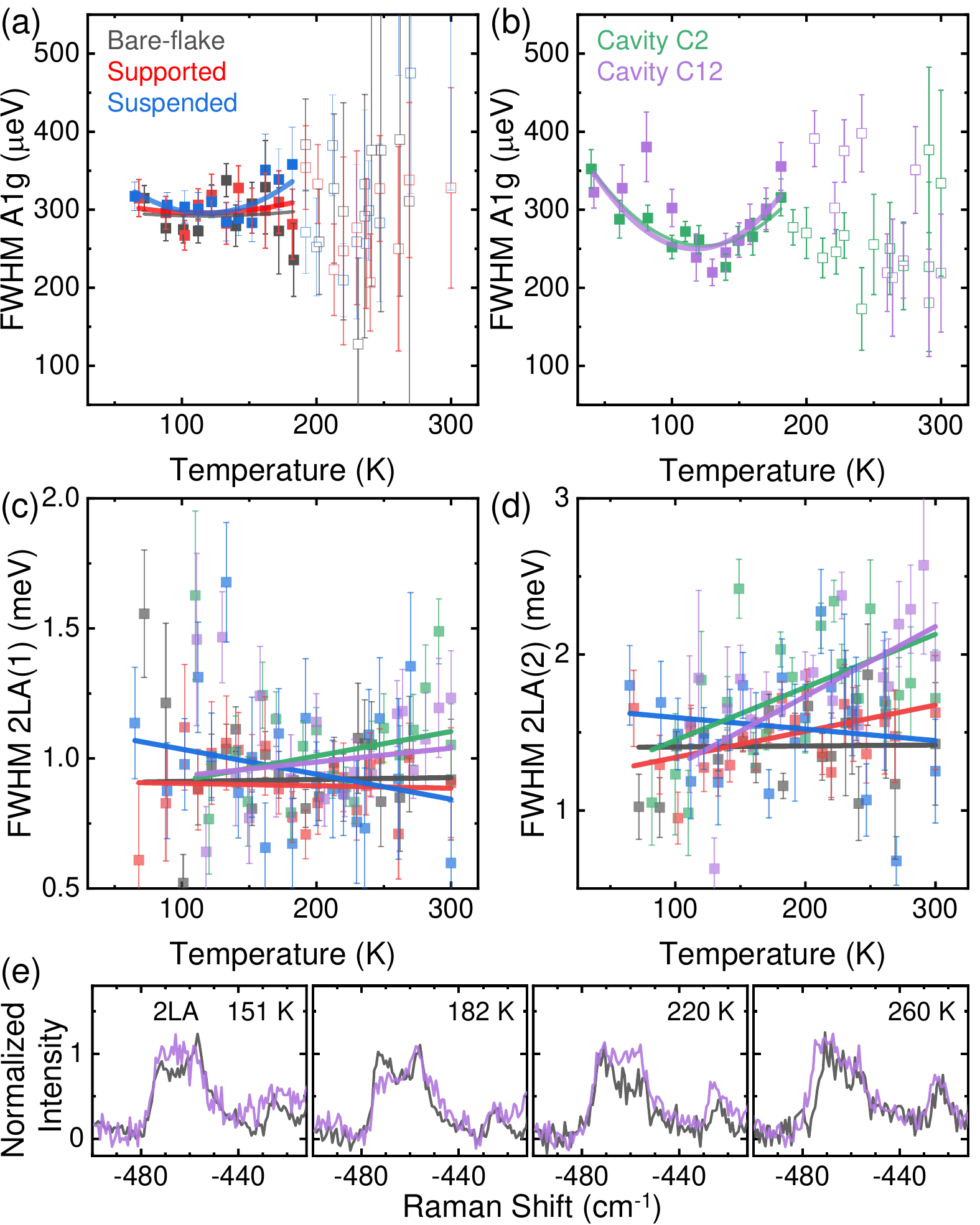}
	\caption{\label{sf11}
		Anharmonicity in Raman linewidths.
		(a)(b) A$_{1g}$ linewidth in (a) control cases and (b) cavities.
		Solid lines are the quadratic fittings.
		Hollow points are data with low SNR.
		(c)(d) Linewidths of two 2LA peaks (SFig.~\ref{sf2}(b)) and the linear fittings (solid lines).
		(f) Raw Raman spectra normalized by the 2LA peak for comparison.
	}
\end{figure}

The tripartite exciton-phonon-phonon coupling will result in an anharmonicity in Raman linewidths \cite{LIU2019451}, which are indeed observed in our experiments as presented in SFig.~\ref{sf11}.
The exciton-phonon coupling introduces a temperature narrowing in the Raman linewidth whilst the phonon-phonon coupling introduces a broadening.
Thus, the Raman linewidth is nonmonotonic at low temperature \cite{LIU2019451}.
Compared to the control cases in SFig.~\ref{sf11}(a), we observe this nonmonotonicity in the A$_{1g}$ linewidth in the cavities presented in SFig.~\ref{sf11}(b).
We note that for the A$_{1g}$ linewidth in SFig.~\ref{sf11}(a)(b), data points at high $T>170\ \mathrm{K}$ (hollow points) has huge error bar due to the low signal-to-noise ratio (SNR) of A$_{1g}$ peak in spectra.
The linewidth of two 2LA peaks extracted by the multi Lorentz fitting (SFig.~\ref{sf2}) are presented in SFig.~\ref{sf11}(c)(d).
Since the data is a bit noisy, here we simply use a linear fitting to quantify the temperature dependence.
As shown, for both two 2LA peaks the broadening in cavities are larger than that in control cases.
We plot the comparison of raw data in SFig.~\ref{sf11}(e) to show this anharmonicity straightforwardly.
The anharmonicity in Raman linewidth further strengthens the tripartite coupling.

We note that the linear fitting in SFig.~\ref{sf11}(c)(d) is a simple method to provide a quantitative comparison of the temperature dependence.
The broadening of Raman linewidth from the phonon-phonon coupling can be generally expressed by a polynomial equation of the Bose factor, and the polynomial order is related to the number of phonons involved in the coupling \cite{LIU2019451}.
In our sample, the Bose factor of cavity vibrational phonon is $\approx  k_B T/\left(\hbar\omega_{cP}\right)$ due to the low energies $\hbar\omega_{cP}\ll k_B T$ as discussed in Sec.~\ref{secs2}.
Therefore, we can expect the Raman linewidth in the cavity following a polynomial equation $\sum c_jT^j$.
The number of cavity vibrational phonons involved in the coupling determines $j$, whilst the frequency of cavity vibrational phonon is reflected in the coefficient $c_j$.
In cavity C12 two cavity phonons are involved in the tripartite coupling (Fig,~3(a)), thereby the Raman linewidth is expected to have a quadratic temperature dependence.
We have tried the quadratic fitting for the data from cavity C12 in SFig.~\ref{sf11}(d), but the improvement is limited by the noise.
Nonetheless, the scope here is that the anharmonicity is observed and further strengthens the tripartite coupling.
Specific details in the anharmonicity could be an interesting future topic.

\begin{figure}
	\includegraphics[width=\linewidth]{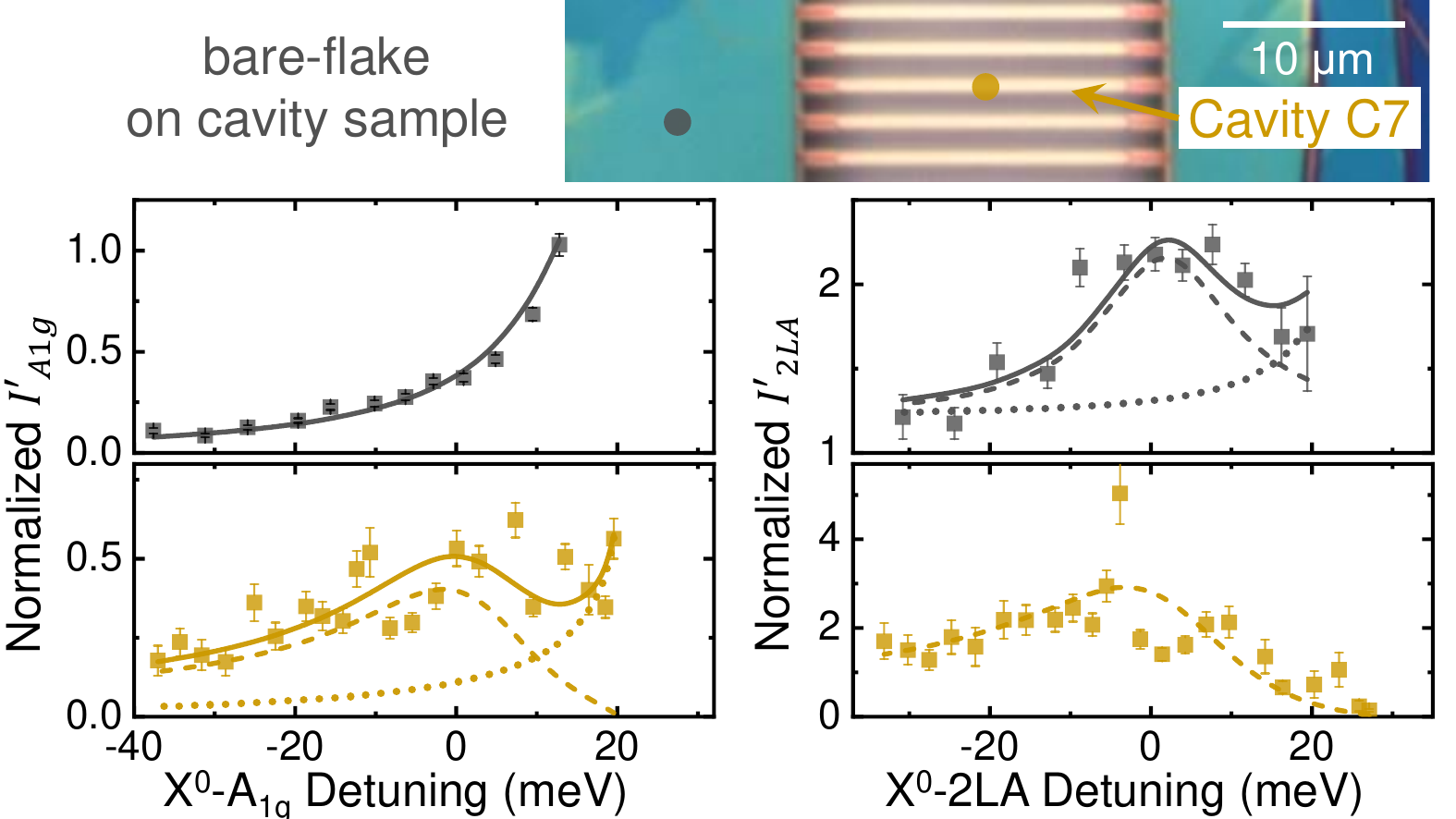}
	\caption{\label{sf12}
		Additional Raman results to show the generality.
		The bare flake on the cavity sample (gray) exhibits same to the bare flake on the control experiment sample in Fig.~2.
		The cavity C7 (yellow) exhibits same to cavity C2 and C12.
	}
\end{figure}

The 2D flakes for the control experiment sample and cavity sample in this work are from same bulk materials and prepared by the same methods discussed in Sec.~\ref{secs1a}.
Nevertheless, one might wonder that the differences between cavities and control cases are from uncertainties during the fabrication.
Here in SFig.~\ref{sf12} we show the Raman datasets measured from the bare flake on the cavity sample (gray) and another cavity C7 (yellow).
As shown, the key results of the bare flake on the cavity sample (gray), which means the observation of X$^-$-A$_{1g}$ peak for A$_{1g}$ whilst both X$^0$-2LA and X$^-$-2LA peaks for 2LA, are same to the bare flake on the control experiment sample in Fig.~2.
The key results of cavity C7 (yellow) are also same to those of C2 and C12, which means the observation of both X$^-$-A$_{1g}$ and X$^0$-A$_{1g}$ peaks for A$_{1g}$ whilst only X$^0$-2LA peak for 2LA.
These results exclude the uncertainties during the sample fabrication.
In addition, the enhancement of $g_{\mathrm{X}^0,\mathrm{2LA}}$ in cavity C7 exhibits a $T^3$ dependence ($N=2.79 \pm 0.47$) that further strengthens a discrete number of cavity phonons participate in the tripartite exciton-phonon-phonon coupling.

\section{\label{secs4}PL Spectroscopy}

\subsection{\label{secs4a}Selective Exciton--Cavity-Phonon Coupling}

\begin{figure}
	\includegraphics[width=\linewidth]{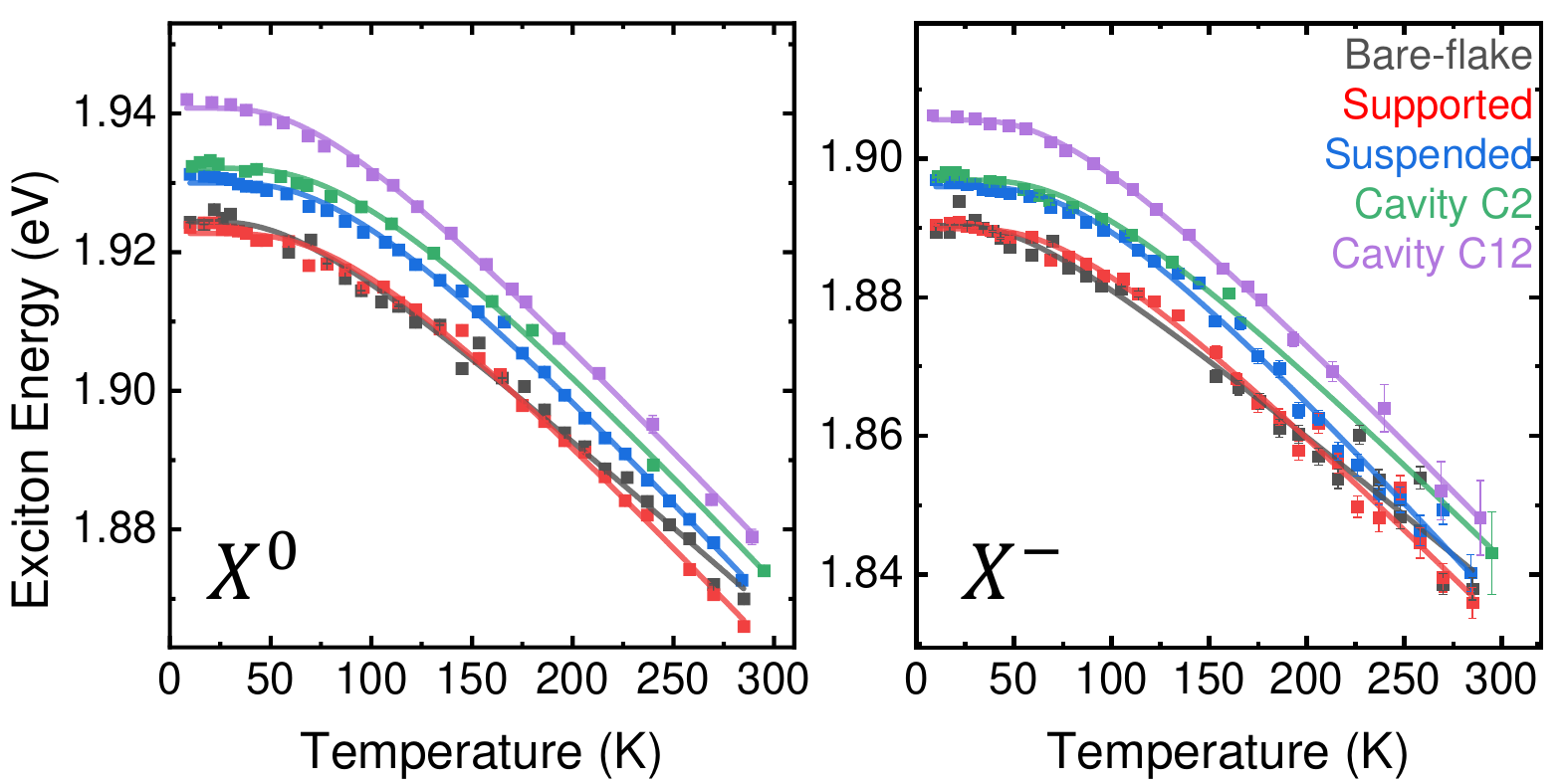}
	\caption{\label{sf13}
		Temperature-dependent energy shift of X$^0$ and X$^-$ in five cases.
		All follow the standard relation.
	}
\end{figure}

We present the emission peak energies of X$^0$ and X$^-$ extracted from PL spectroscopy in SFig.~\ref{sf13}.
The exciton energy in TMDs usually follows the standard hyperbolic cotangent relation \cite{PhysRevX.7.021026} as
\begin{eqnarray}
	\label{eqe}
	\omega_X=\omega_0-S\langle\hbar\omega\rangle_E\lbrack\coth\left({\frac{\langle\hbar\omega\rangle_E}{k_B T}}\right)-1\rbrack
\end{eqnarray}
where $\omega_0$ is the exciton energy at zero temperature, $S$ is a dimensionless coupling constant, $\langle\hbar\omega\rangle_E$ is an average phonon energy and $k_B$ is the Boltzmann constant.
In both control cases and cavities, both the X$^0$ and X$^-$ energy is well described by Eq.~(\ref{eqe}).
For X$^0$ in the bare flake, supported, suspended, cavity C2 and cavity C12 cases, we observe $S=1.52\pm0.06,\ 1.87\pm0.06,\ 1.89\pm0.04,\ 1.88\pm0.06,\ 1.82\pm0.04$ and $\langle\hbar\omega\rangle_E=16\pm2,\ 23\pm1,\ 22\pm1,\ 24\pm1,\ 18\pm1$ meV, respectively.
For X$^-$, we observe $S=1.37\pm0.10,\ 1.69\pm0.08,\ 1.88\pm0.08,\ 1.68\pm0.07,\ 1.75\pm0.04$ and $\langle\hbar\omega\rangle_E=14\pm3,\ 21\pm2,\ 23\pm2,\ 22\pm2,\ 19\pm1$ meV respectively.
Little difference is observed between the five cases, and all fitting parameters agree well to previous reports \cite{PhysRevX.7.021026}.

As discussed in the main paper, since in Raman spectra (excited by 632 nm-laser) the exciton emission line shape is strongly deformed by the phonon couplings \cite{Molas2017}, exciton emission properties directly extracted from Raman spectra might be inaccurate.
Therefore, we calculate the exciton emission properties in Raman spectra by the PL data presented here.
We use Eq.~(\ref{eqe}), with $S$ and $\langle\hbar\omega\rangle_E$ extracted from SFig.~\ref{sf13}, in addition with an $\omega_0'$ aligned by the X$^0$-Raman resonant $T$ point, to calculate the X$^0$ and X$^-$ energies in the Raman analyses.
We have to align $\omega_0'$ since the different measurement setups for PL and Raman spectroscopy introduce inaccuracy in the absolute value of photon energies in the spectra.
The X$^0$ and X$^-$ linewidths are calculated by the fitting curves in SFig.~\ref{sf16} discussed later.

\begin{figure}
	\includegraphics[width=\linewidth]{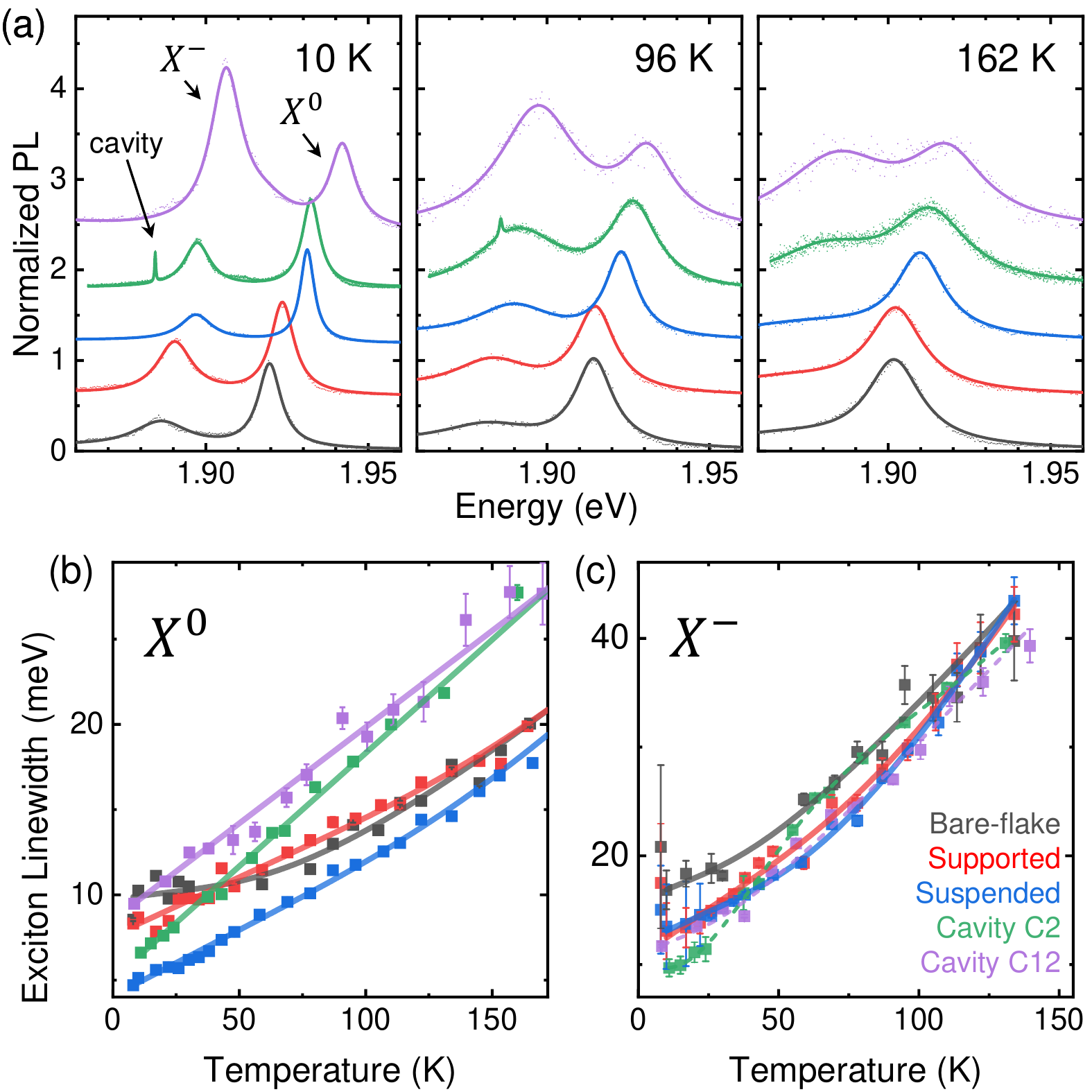}
	\caption{\label{sf14}
		(a) Typical PL spectra recorded from five cases around 10, 96 and 162 K.
		The sharp peak in cavity C2 at 10 K is from the photonic mode.
		(b) Temperature-dependent exciton linewidth of X$^0$ and (c) X$^-$ in five cases.
	}
\end{figure}

We present the emission peak linewidth of X$^0$ and X$^-$ at low temperature $T<175\ \mathrm{K}$ in SFig.~\ref{sf14}.
Typical PL spectra recorded from five cases around 10, 96 and 162 K are presented in SFig.~\ref{sf14}(a) for comparison.
The exciton linewidths at low $T$ are nearly same in five cases.
However, as $T$ increases, the X$^0$ linewidth in the cavities exhibits a significant broadening compared to that in control cases.
This broadening indicates the exciton-phonon coupling is modulated in the cavities.

The exciton linewidth in 2D semiconductors usually follows the phenomenological equation \cite{PhysRevX.7.021026,Selig2016,PhysRevLett.116.127402}
\begin{eqnarray}
	\label{eql}
	\gamma=\gamma_0+a_1 T+\frac{a_2}{\mathrm{exp}\left(\frac{\langle\hbar\omega\rangle_L}{k_B T}\right)-1}
\end{eqnarray}
where $\gamma_0$ is the intrinsic X$^0$ or X$^-$ linewidth, $a_1$ (slope at low $T$) quantifies the strength of linear $T$-broadening induced by low-energy acoustic phonons and $a_2$ is the nonlinear broadening arising from high-energy phonons with an average energy $\langle\hbar\omega\rangle_L$.
Measured linewidths of X$^0$ ($\gamma_{X^0}$) and X$^-$ ($\gamma_{X^-}$) are presented in SFig.~\ref{sf14}(b)(c) respectively.
For $\gamma_{X^0}$ in SFig.~\ref{sf14}(b), the three control cases (gray, red, blue) are well described by Eq.~(\ref{eql}).
The fitting terms in the bare flake case (gray) quantitatively agree with previous reports, including the coupling strengths $a_{1,\mathrm{X}^0}=18.5\pm 1.7\ \mathrm{\mu eV\cdot K^{-1}}$ and $a_{2,X^0}=29.8\pm 5.8\ \mathrm{meV}$ \cite{Selig2016,PhysRevX.7.021026,PhysRevLett.116.127402}.
In the supported (red) and suspended (blue) cases, local static strain from the Si$_3$N$_4$ and the suspension is induced as discussed in the main paper.
The best fit values $a_{1,\mathrm{X}^0}=68.5\pm 4.9\ (77.5\pm 3.2)\ \mathrm{\mu eV\cdot K^{-1}}$ and $a_{2,X^0}=91\pm 48\ (77\pm 20)\ \mathrm{meV}$ for the supported (suspended) case are larger than those in the bare flake, consistent with strain-induced cases reported previously \cite{doi:10.1063/1.4985299,C8NR00332G,Khatibi_2018}.
In contrast to three control cases, $\gamma_{X^0}$ recorded from the cavities behaves quite differently.
Specifically, we observe a linear $T$-broadening up to $150\ \mathrm{K}$, with the slope $a_{1,\mathrm{X}^0}=133\pm 2\ (113\pm 4)\ \mathrm{\mu eV\cdot K^{-1}}$ in cavity C2 (C12), almost double the value obtained in the control cases.
This enhanced value of $a_{1,\mathrm{X}^0}$ indicates that the coupling between X$^0$ and low-energy phonons in the cavity is stronger compared to the control cases, unveiling the contribution from cavity vibrational phonons.
The cavity phonons exactly have low energies $\ll k_B T$ as discussed in Sec.~\ref{secs1a} thereby contributes to the linear broadening \cite{Selig2016,PhysRevX.7.021026}.

\begin{figure*}
	\includegraphics[width=0.8\linewidth]{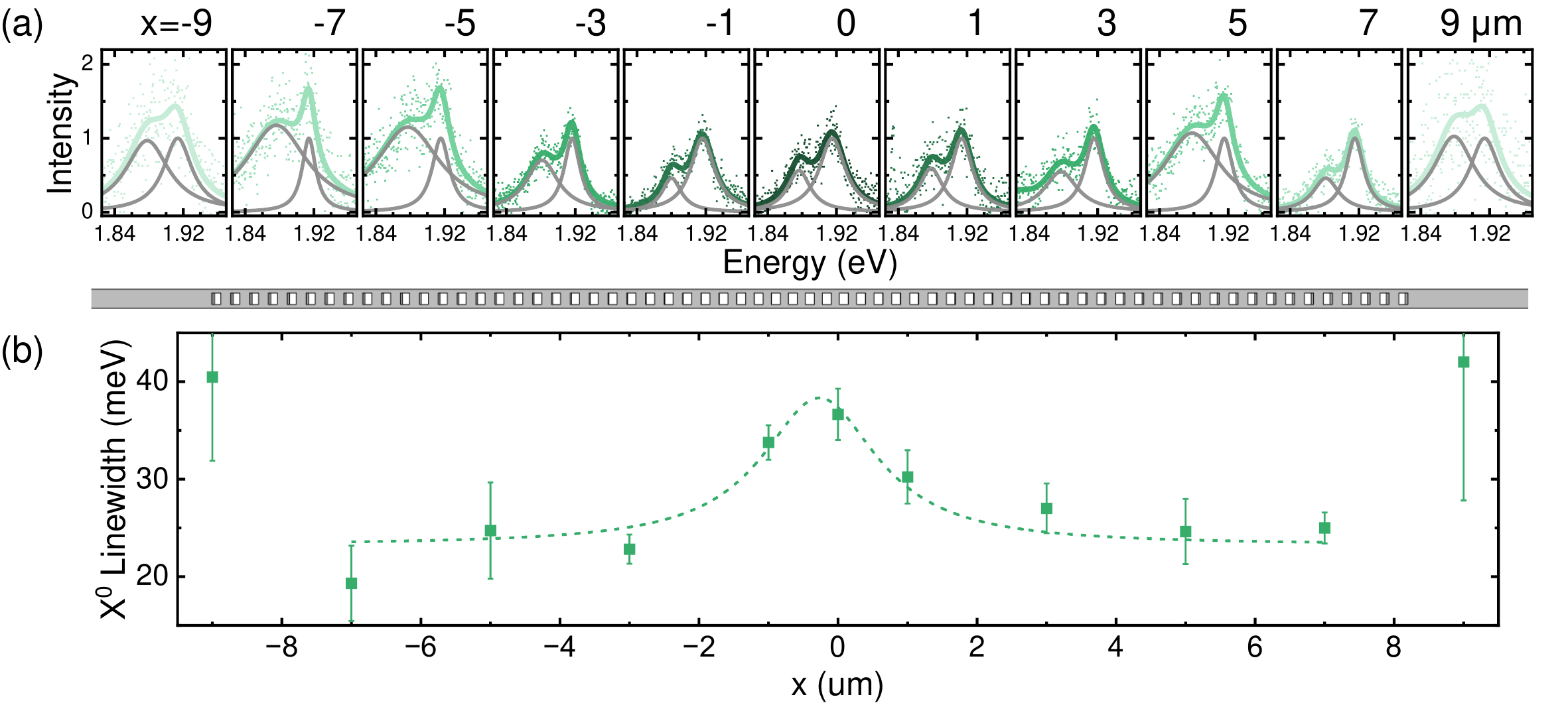}
	\caption{\label{sf15}
		Position dependence of X$^0$ linewidth broadening.
		(a) Position-dependent PL spectra of cavity C2 measured at 170 K, normalized by X$^0$ peak at 1.91 eV.
		(b) Extracted X$^0$ linewidth along the nanobeam, exhibiting a centralized broadening.
	}
\end{figure*}

In contrast to X$^0$, the X$^-$ linewidth shows little difference between the four cases as presented in SFig.~\ref{sf14}(c).
The inaccuracy of $\gamma_{X^-}$ at low $T$ is mainly from the fine structure of MoS$_2$ trions \cite{PhysRevB.105.L041302} shown in SFig.~\ref{sf3}(b).
Despite this, $\gamma_{X^-}$ is well described by Eq.~(\ref{eql}) for the three control experiments (gray, red, blue), and the linear term $a_{1,\mathrm{X}^-}$ is $106\pm 31,\ 154\pm 20,\ 114\pm 8 \ \mathrm{\mu eV\cdot K^{-1}}$, respectively.
Although $\gamma_{X^-}$ in cavities (green, purple) cannot be appropriately described by Eq.~(\ref{eql}), no additional broadening beyond the control cases is observed.
Based on the PL linewidths in SFig.~\ref{sf14}, we conclude that the X$^0$--cavity-phonon coupling strength is much stronger than the X$^-$--cavity-phonon coupling strength.
This selectivity in the bipartite coupling between excitons and cavity phonons perfectly agrees with the selectivity in the tripartite coupling.
Similar phenomena that different excitons couple to different phonons have also been observed in other semiconductors \cite{doi.org/10.1002/lpor.201900267}.

In addition, the coupling strength $a_{1,\mathrm{X}^0}$ and $a_{1,\mathrm{X}^-}$ extracted from PL spectra provide a comparison to the coupling strength $g_{\mathrm{X}^0,\mathrm{2LA}}$ and $g_{\mathrm{X}^-,\mathrm{2LA}}$ in the fitting of Raman spectra $I'_{\mathrm{2LA}}=g_{\mathrm{X}^0,\mathrm{2LA}}R_{\mathrm{X}^0,\mathrm{2LA}}+g_{\mathrm{X}^-,\mathrm{2LA}}R_{\mathrm{X}^-,\mathrm{2LA}}$ presented in Fig.~2(a).
The ratio $a_{1,\mathrm{X}^-}/a_{1,\mathrm{X}^0}$ is $5.7\pm 2.2,\ 2.2\pm 0.5,\ 1.5\pm 0.2$ for the bare flake, supported and suspended case, respectively.
The ratio $g_{\mathrm{X}^0,\mathrm{2LA}}/g_{\mathrm{X}^-,\mathrm{2LA}}$ in the three control cases $5.1\pm 1.6,\ 2.0\pm 0.6,\ 2.0\pm 0.4$ generally agree to those observed from the PL spectra.
This agreement further supports our fitting method discussed for SFig.~\ref{sf8}.

To further strengthen the X$^0$--cavity-phonon coupling, we record the position-dependent PL spectra along cavity C2, as presented in SFig.~\ref{sf15}(a).
The X$^0$ linewidth $\gamma_{X^0}$ is presented in SFig.~\ref{sf15}(b).
As expected, $\gamma_{X^0}$ exhibits a centralized broadening due to the cavity vibrational modes are centralized (SFig.~\ref{sf10}), further supporting the X$^0$--cavity-phonon coupling discussed in SFig.~\ref{sf14}.
These agreements between Raman and PL spectroscopy, including the selectivity to X$^0$ (Fig.~2 and SFig.~\ref{sf14}) and the position dependence (SFig.~\ref{sf10} and SFig.~\ref{sf15}), further strengthens the conclusions of phononic hybridization.

In contrast to the position dependence of Raman enhancement presented in SFig.~\ref{sf10}(b), the position dependence in PL linewidth is less clear, e.g., we cannot distinguish the contributions from beam and confined modes in SFig.~\ref{sf15}(b).
The worse distinguishability in PL datasets compared to Raman datasets is since that Raman spectroscopy is the direct measurement of phonons.
In contrast, PL spectroscopy is an indirect measurement i.e., Eq.~(\ref{eql}) shows that the exciton-phonon coupling is reflected by the $T$-broadening rather not absolute value of exciton linewidth.
Besides the exciton-phonon coupling, there exist other factors affecting the exciton linewidth such as the quality of TMD (commercial or grown) \cite{PhysRevX.7.021026}.
As a result, usually $\gamma_{X^0}$ is slightly different in different cases, which is clearly shown in SFig.~\ref{sf14}.
Nonetheless, the position dependence of $\gamma_{X^0}$ in SFig.~\ref{sf15} clearly exhibits the centralized broadening, indicating the exciton-phonon coupling dominates the exciton linewidth as the temperature increases.

\subsection{\label{secs4b}Phonon-Phonon Nonlinearity}

In Sec.~\ref{secs4a} we mainly focus on the exciton linewidth at low temperature below 175 K.
Before presenting the PL data at high temperature, we first discuss the limitation of PL datasets.

In SFig.~\ref{sf14}, we use phenomenological equation Eq.~(\ref{eql}) to fit the exciton linewidth in control cases (gray, red, blue) and use the linear broadening $\gamma=\gamma_0+a_1 T$ to fit the X$^0$ linewidth below 175 K in the cavities.
The linear term $a_1 T$ has the clear physical meaning -- coupling to low-energy phonons -- because for phonon with low energy $\hbar\omega$, the Bose factor $\approx {k_BT}/\left({\hbar\omega}\right)\propto T$ as discussed in Sec.~\ref{secs2}, and the corresponding broadening of exciton linewidth is also $\propto T$ \cite{Selig2016}.
Thus, the broadening from different low-energy phonons can be summed linearly, and the sum is still proportional to T such as
\begin{eqnarray}
	\label{suml}
	c_1\frac{k_BT}{\hbar\omega_1}+c_2\frac{k_BT}{\hbar\omega_2}=\langle c \rangle\frac{k_BT}{\langle\hbar\omega\rangle}
\end{eqnarray}
where $c_1,\ c_2$ reflects the exciton-phonon coupling with a single phonon and $\langle c \rangle,\ \langle\hbar\omega\rangle$ reflects the average effect of two phonons $\hbar\omega_1$ and $\hbar\omega_2$.
However, the nonlinear term in Eq.~(\ref{eql}) is a phenomenological term with the Bose factor of “average” phonon energy $\langle\hbar\omega\rangle_L$.
But for high-energy phonons, their Bose factor is not proportional to $T$.
The sum of Bose factors of two high-energy phonons cannot be described by a new Bose factor as
\begin{eqnarray}
	\label{sumh}
	\frac{c_1}{\mathrm{exp}\left(\frac{\hbar\omega_1}{k_BT}\right)-1}+\frac{c_2}{\mathrm{exp}\left(\frac{\hbar\omega_2}{k_BT}\right)-1}\neq\frac{\langle c \rangle}{\mathrm{exp}\left(\frac{\langle\hbar\omega\rangle}{k_BT}\right)-1}
\end{eqnarray}
. Thus, the exciton linewidth broadening from different high-energy phonons cannot be simply averaged, and the nonlinear term in Eq.~(\ref{eql}) is not strictly correct.
This is the reason why in Sec.~\ref{secs4a} we mainly focus on PL data below 175 K, where the linear exciton linewidth broadening has direct physical meanings (coupling with low-energy phonons).

\begin{figure}
	\includegraphics[width=\linewidth]{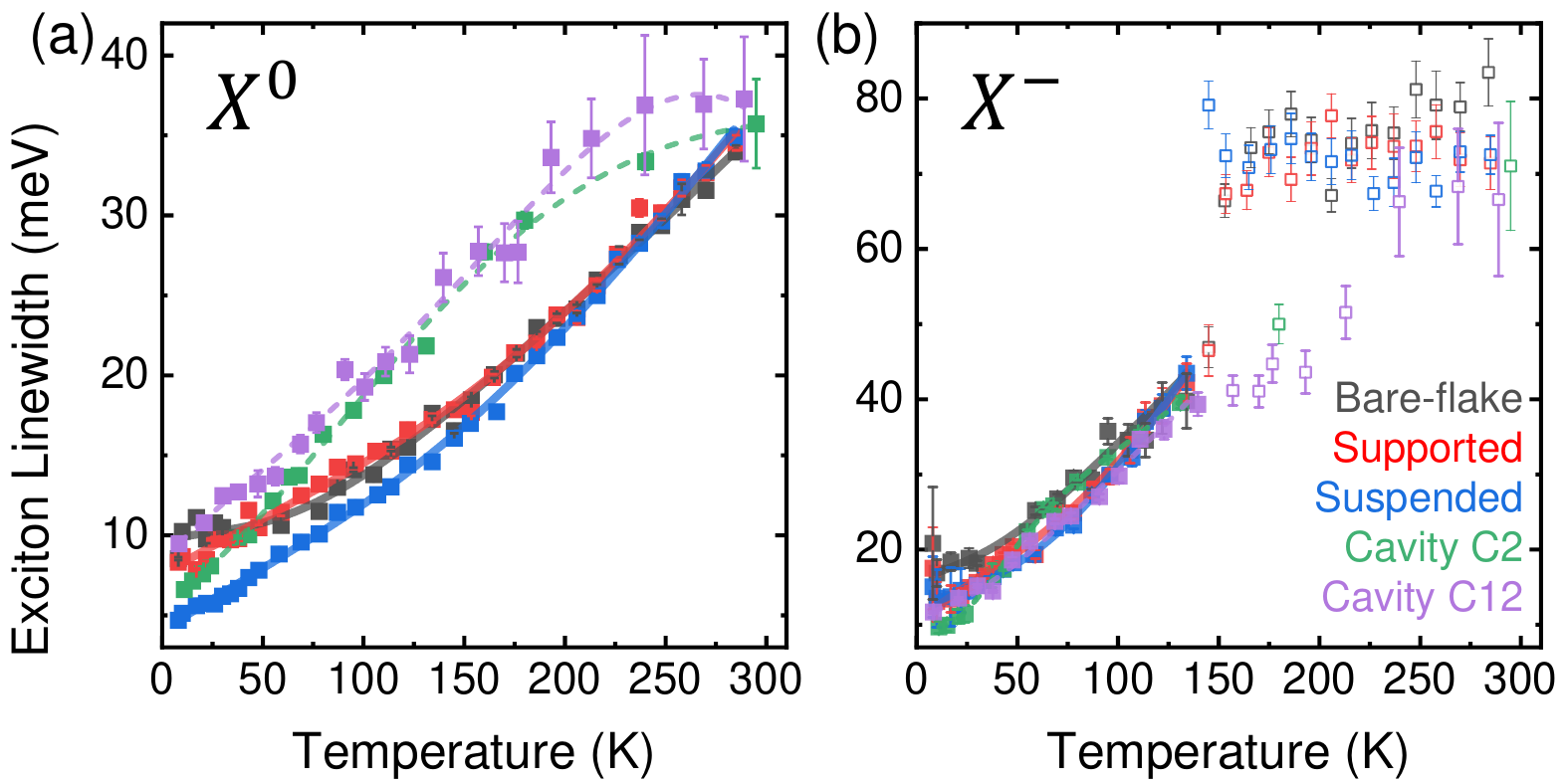}
	\caption{\label{sf16}
		Temperature-dependent linewidth of (a) X$^0$ and (b) X$^-$ in five cases.
		The hollow points in (b) is inaccurate due to the merge of X$^-$ and D$^-$ peak shown in SFig.~\ref{sf3}(b).
	}
\end{figure}

Nonetheless, here we try to provide an explanation of the non-trivial exciton linewidth at high temperature.
We present the X$^0$ and X$^-$ linewidth at all temperature points in SFig.~\ref{sf16}(a)(b), respectively.
In the two cavities (green, purple), the X$^0$ linewidth $\gamma_{X^0}$ in the whole measurement range $T \in (0,\ 300)\ \mathrm{K}$ cannot be described by Eq.~(\ref{eql}), thus is fitted by a polynomial equation
\begin{eqnarray}
	\label{suma}
	\gamma_{X^0}=\gamma_0+c_1T+c_2T^2+c_3T^3+c_4T^4
\end{eqnarray}
shown by the dashed lines in SFig.~\ref{sf3}(a).
We use Eq.~(\ref{suma}) just to sketch a fitting to estimate the value of $\gamma_{X^0}$ at other $T$ points.
The nonlinear parameters $c_i\ \left(i\geq2\right)$ have no general physical meanings as discussed in Eq.~(\ref{sumh}).
Generally, $\gamma_{X^0}$ exhibits similar behavior in both two cavities: approximately a linear $T$-broadening below 175 K whilst the $T$-broadening rate saturates above 175 K.
The non-trivial $\gamma_{X^0}$ at high temperature $T>175\ \mathrm{K}$ further indicates the nonlinear term in Eq.~(\ref{eql}) is not suitable for all cases.

The X$^-$ linewidth $\gamma_{X^-}$ at high temperature $T>150\ \mathrm{K}$ (hollow points in SFig.~\ref{sf16}(b)) is inaccurate due to the overlap between X$^-$ and D$^-$ peak as discussed in SFig.~\ref{sf3}(b).
Moreover, around 150 K where $\gamma_{X^0}$ exhibits maximum difference between cavities and control cases, $\gamma_{X^-}$ exhibits little difference between five cases.
Thereby, the conclusion of weak X$^-$--cavity-phonon coupling is same to that obtained from Raman spectra.
Similar to $\gamma_{X^0}$, we fit $\gamma_{X^-}$ in two cavities by Eq.~(\ref{suma}) just to sketch a fitting to estimate the value of $\gamma_{X^-}$ at other $T$ points.

\begin{figure}
	\includegraphics[width=\linewidth]{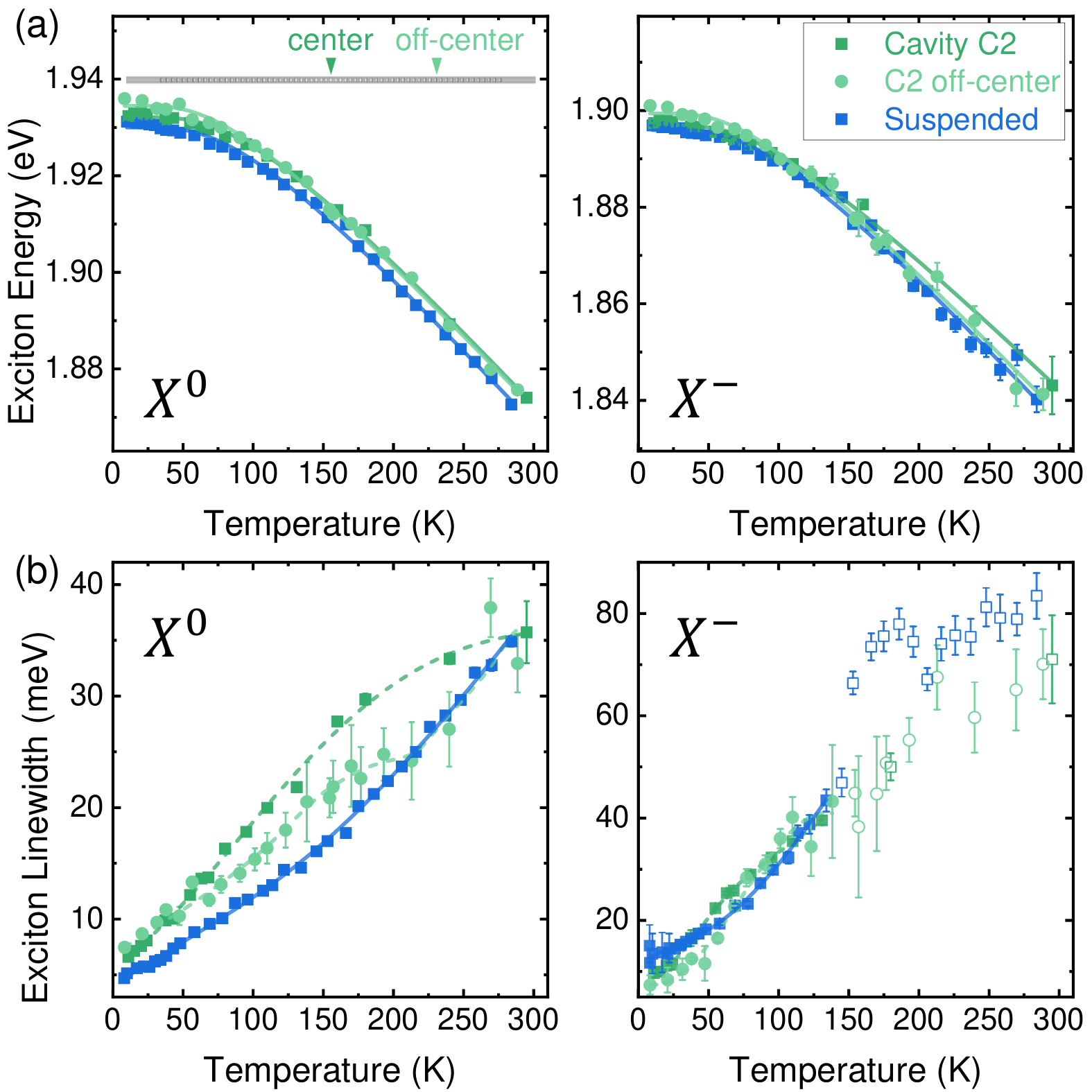}
	\caption{\label{sf17}
		Temperature-dependent (a) exciton energies (b) linewidths measured at cavity C2 off-center (light green dots), $\sim 5\ \mathrm{\mu m}$ away from the nanobeam center position.
		Datasets of cavity C2 center (green rectangles) and suspended case (blue rectangles) from SFig.~\ref{sf13} and SFig.~\ref{sf16} are also plotted here for comparison.
	}
\end{figure}

To further investigate the interplay between the non-trivial $\gamma_{X^0}$ and the cavity vibrational phonons, we next discuss the position dependence of $\gamma_{X^0}$ since the cavity phonons are position-dependent as discussed in SFig.~\ref{sf10} and SFig.~\ref{sf15}.
The absolute value of $\gamma_{X^0}$ in SFig.~\ref{sf15} is not a strict reflection of the exciton-phonon coupling.
Therefore, we measure the temperature dependent PL spectra at an off-center position in cavity C2, which is $\sim 5\ \mathrm{\mu m}$ away from the nanobeam center.
We present the extracted exciton energies and linewidths (light green dots) in SFig.~\ref{sf17}, along with the results from cavity C2 center (green rectangles) and the suspended case (blue rectangles) for comparison.
For the exciton energies in the off-center case (SFig.~\ref{sf17}(a)), we observe $S=1.83\pm0.05,\ \langle\hbar\omega\rangle_E=20\pm1$ meV for X$^0$ and $S=1.78\pm0.08,\ \langle\hbar\omega\rangle_E=19\pm2$ meV for X$^-$.
These values are same to those observed in other cases (SFig.~\ref{sf13}) and in previous works \cite{PhysRevX.7.021026}, thereby further strengthen the conclusion that the exciton energies are rarely affected by the cavity vibrational phonons.

The exciton linewidths in the off-center case presented in SFig.~\ref{sf17}(b) generally exhibit similar behavior to those at center.
Same to SFig.~\ref{sf14}(c) and SFig.~\ref{sf16}(b), the X$^-$ linewidth $\gamma_{X^-}$ exhibits little difference between the center, off-center and control cases.
This agreement further strengthens the weak X$^-$--cavity-phonon coupling.
As $T$ increases, the X$^0$ linewidth $\gamma_{X^0}$ in the off-center case firstly behavior like that at center: approximately a linear $T$-broadening below 120 K and then a saturation over 120 K.
However, after being “caught up” by the suspended case at 210 K, the $T$-broadening rate of $\gamma_{X^0}$ in the off-center case again increases.
At $T>210\ \mathrm{K}$, $\gamma_{X^0}$ in the off-center case approximately follows the suspended case.
Generally, the off-center case behavior like a “medium” between the cavity center and the control case.
The “medium” is reasonable, since the off-center case has part of the cavity vibrational phonons (beam modes) but not all (no confined modes) as discussed in SFig.~\ref{sf10}.

Finally, we try to explain the non-trivial X$^0$ linewidth at high temperature.
One possible reason is the nonlinearity in mechanics, i.e., nonlinearity in the Young's moduli and deformation potentials \cite{DU202133525,PhysRevB.19.2209}.
The mechanics discussed up to now is in linear regime which means the deformation-induced strain and electron (hole) potential change are all proportional to the deformation.
In this linear regime, the vibration amplitude (cavity phonon population) does not affect the phononic mode (frequency and $Q$ factor). Generally, the mechanics is linear with small deformations but nonlinear with large deformations.
E.g., as the cavity vibration amplitude (phonon population) increases with $T$, we could image above some points the strain increases superlinearly to the deformation, thus the vibration saturates.
However, the nonlinearity in cavity vibration is usually observed with strong external driving such as a piezo with huge AC voltage \cite{doi:10.1021/acs.nanolett.0c01586}.
The natural thermal excited vibrations usually have small deformations within the linear regime.
Thereby, we think the nonlinearity in mechanics is improbable.

As such, we suggest the saturation in $T$-broadening is not due to the nonlinearity of mechanics, but rather due to the nonlinearity in the broadening effects of the phonon couplings.
Usually, the decay rate (linewidth) of an exciton $\gamma_X$ is written as
\begin{eqnarray}
	\label{dec}
	\gamma_X=\gamma_{rad}+\gamma_{non-rad}
\end{eqnarray}
where $\gamma_{rad}$ is the radiative decay rate and $\gamma_{non-rad}$ is the non-radiative decay and dephasing rate written as
\begin{eqnarray}
	\label{dec1}
	\gamma_{non-rad}=\sum_i\gamma_{phi}+\gamma_{others}
\end{eqnarray}
where $\gamma_{phi},\ i=1,2,\cdots$ means the exciton-phonon coupling to phonon ph1, ph2, $\cdots$ and $\gamma_{others}$ describes other non-radiative decay channels.
We emphasize that here the decay rate $\gamma_{rad}$, $\gamma_{ph1}$, $\gamma_{ph2}$, $\cdots$ and $\gamma_{others}$ are summed linearly.
This linear sum means these decay channels are independent to each other.
The independence is usually valid, e.g., the phonon-phonon coupling between different lattice phonons is usually very weak, thus $\gamma_{ph1}+\gamma_{ph2}$ is reasonable when ph1 and ph2 are both lattice phonons.

However, the cavity phonon is not independent to the lattice phonon.
We observe the lattice-phonon—cavity-phonon coupling, and such coupling increases with temperature $T$.
Therefore, we could expect that the sum of $\gamma_{cavity}$ (broadening from cavity phonons) and $\gamma_{lattice}$ (broadening from lattice phonons) is approximately linear (independent) at low $T$ as $\gamma_{cavity}+\gamma_{lattice}$ but nonlinear (correlated) at high $T$ such as
\begin{eqnarray}
	\label{dec4}
	\left(\gamma_{cavity}^k+\gamma_{lattice}^k\right)^{1/k}
\end{eqnarray}
where $k$ is a dimensionless quantitation of the phonon-phonon nonlinearity (coupling) increasing with $T$.
Therefore, at high $T$ the phonon-induced linewidth broadening is smaller than $\gamma_{cavity}+\gamma_{lattice}$, resulting in the saturation we observed in experiments.

Based on this nonlinearity between cavity and lattice phonons, we propose a phenomenological model to explain X$^0$ linewidth.
The lattice-phonon-induced broadening is written as
\begin{eqnarray}
	\label{decl}
	\gamma_{lattice}=a_1 T+\frac{a_2}{\mathrm{exp}\left(\frac{\langle\hbar\omega\rangle_L}{k_B T}\right)-1}
\end{eqnarray}
following the Eq.~(\ref{eql}).
The cavity-phonon-induced broadening is written as
\begin{eqnarray}
	\label{decc}
	\gamma_{cavity}=c_1 T
\end{eqnarray}
since the cavity phonons have low energies as discussed in Sec.~\ref{secs1a}.
The phonon-induced broadening is then calculated by Eq.~(\ref{dec4}) as
\begin{eqnarray}
	\label{decf}
	\gamma_{phonon}=\left(\gamma_{cavity}^k+\gamma_{lattice}^k\right)^{1/k}
\end{eqnarray}
. We set $k$ which describes the phonon-phonon nonlinearity (coupling) as
\begin{eqnarray}
	\label{deck}
	k=1+\left(\frac{T}{T_0}\right)^m
\end{eqnarray}
where $T_0$ describes the saturation temperature and $m$ is a dimensionless parameter.

\begin{figure}
	\includegraphics[width=\linewidth]{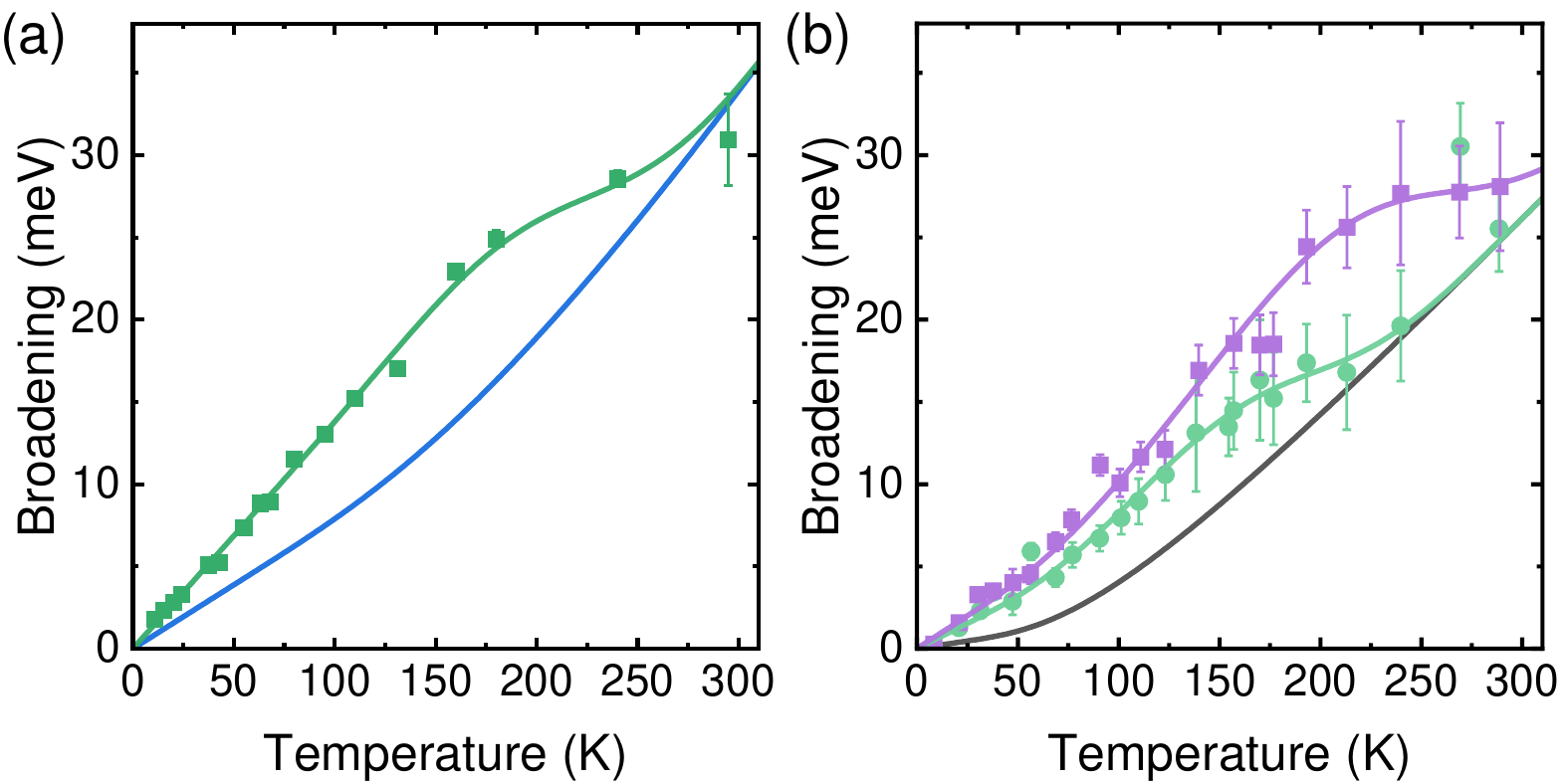}
	\caption{\label{sf18}
		Fitting of non-trivial X$^0$ linewidth based on the nonlinearity between lattice and cavity phonons.
		(a) Cavity C2 (green) based on $\gamma_{lattice}$ from the suspended case (blue).
		(b) Cavity C2 off-center (light green) and cavity C12 (purple) based on $\gamma_{lattice}$ from the bare flake (gray).
	}
\end{figure}

We present the fitting of cavity C2 based on Eq.~(\ref{decf}) in SFig.~\ref{sf18}(a). The intrinsic X$^0$ linewidth $\gamma_0$ at zero temperature is $5.0\pm0.1$ meV in cavity C2, close to that value in the suspended case $4.1\pm0.2$ meV.
Thereby, we use $\gamma_{lattice}$ extracted from the suspended case (blue line) for the fitting, including $a_{1,\mathrm{X}^0}=77.5\pm 3.2\ \mathrm{\mu eV\cdot K^{-1}}$, $a_{2,X^0}=77\pm 20\ \mathrm{meV}$ and $\langle\hbar\omega\rangle_L=54.6\pm 6.8\ \mathrm{meV}$.
The green line in SFig.~\ref{sf18}(a) is the fitting calculated with $c_1=60\pm3\ \mathrm{\mu eV/K}$ for $\gamma_{cavity}$, $T_0=238\pm9\ \mathrm{K}$ and $m=6$.
In contrast, $\gamma_0$ in cavity C2 off-center and cavity C12 is $6.9\pm0.4$ and $8.5\pm0.2$ meV respectively, close to that value in the bare flake case $9.7\pm0.4$ meV.
Thereby, we fit these two cases based on $\gamma_{lattice}$ extracted from the bare flake case (gray line) as presented in SFig.~\ref{sf18}(b).
Parameters for $\gamma_{lattice}$ are $a_{1,\mathrm{X}^0}=18.5\pm 1.7\ \mathrm{\mu eV\cdot K^{-1}}$, $a_{2,X^0}=29.8\pm 5.8\ \mathrm{meV}$ and $\langle\hbar\omega\rangle_L=23.1\pm 5.8\ \mathrm{meV}$.
The cavity C2 off-center position (light green line) is fitted with $c_1=43\pm3\ \mathrm{\mu eV/K}$, $T_0=205\pm8\ \mathrm{K}$ and $m=6$.
The cavity C12 (purple line) is fitted with $c_1=61\pm2\ \mathrm{\mu eV/K}$, $T_0=270\pm6\ \mathrm{K}$ and $m=7$.
As presented in SFig.~\ref{sf18}, the fittings quantitatively agree with the experimental results.
The X$^0$--cavity-phonon coupling term $c_1$ is smaller at the off-center position compared to the values at two cavity centers, consistent to the discussions in SFig.~\ref{sf10}, SFig.~\ref{sf15} and SFig.~\ref{sf17}.
The saturation temperature $T_0$ is also smaller at the off-center position.
We explain this by the larger population of the beam vibrational phonons off-center (MHz) compared to the confined vibrational phonons at center (GHz).
Therefore, the population required for the nonlinearity corresponds to lower $T_0$ for the beam vibrational phonons (off-center) and higher $T_0$ for the confined vibrational phonons (at center).

We emphasize that the exciton linewidth at high temperature is non-trivial and not deeply studied yet.
We propose the model of Eqs.~(\ref{decc})-(\ref{deck}) to provide a phenomenological explanation of our experimental observations.
The fitting results in SFig.~\ref{sf18} show that the nonlinearity between cavity and lattice phonons described in our model can quantitatively reproduce the experimental results, but we cannot exclude all other possibilities.
Nonetheless, the non-trivial X$^0$ linewidth at high temperature discussed here affect neither our conclusions in the main paper based on Raman spectra, nor the conclusions in Sec.~\ref{secs4a} based on low-temperature PL data.

\end{document}